\documentclass[preprints,article,accept,moreauthors,pdftex]{mdpi}
\firstpage{1} 
\pubvolume{xx}
\issuenum{x}
\articlenumber{y}
\pubyear{2023}
\copyrightyear{2023}
\history{}
\usepackage{subcaption}

\Title{An adaptive safety layer with hard constraints for safe reinforcement learning in multi-energy management systems}

\Author{Glenn Ceusters $^{1,2,3,*}$\orcidA{},  Muhammad Andy Putratama $^{2}$\orcidB{}, Rüdiger Franke $^{1}$, Ann Nowé $^{3}$\orcidC{}, Maarten Messagie $^{2}$\orcidD{}}

\AuthorNames{Glenn Ceusters, Muhammad Andy Putratama, Rüdiger Franke, Ann Nowé, Maarten Messagie}

\address{%
$^{1}$ \quad ABB, Hoge Wei 27, 1930 Zaventem, Belgium; glenn.ceusters@be.abb.com; ruediger.franke@de.abb.com;\\
$^{2}$ \quad Vrije Universiteit Brussel (VUB), ETEC-MOBI, Pleinlaan 2, 1050 Brussels, Belgium; glenn.leo.ceusters@vub.be; muhammad.andy.putratama@vub.be, maarten.messagie@vub.be;\\
$^{3}$ \quad Vrije Universiteit Brussel (VUB), AI-lab, Pleinlaan 2, 1050 Brussels, Belgium; gceusters@ai.vub.ac.be; ann.nowe@ai.vub.ac.be; \\
}

\corres{\textbf{Correspondence:} glenn.ceusters@be.abb.com}

\abstract{Safe reinforcement learning (RL) with hard constraint guarantees is a promising optimal control direction for multi-energy management systems. It only requires the environment-specific constraint functions itself \textit{a priori} and not a complete model (i.e., plant, disturbance and noise models, and prediction models for states not included in the plant model - e.g. demand forecasts, weather forecasts, price forecasts). The project-specific upfront and ongoing engineering efforts are therefore still reduced, better representations of the underlying system dynamics can still be learnt, and modelling bias is kept to a minimum (no model-based objective function). However, even the constraint functions alone are not always trivial to accurately provide in advance (e.g., an energy balance constraint requires the detailed determination of all energy inputs and outputs), leading to potentially unsafe behaviour. Furthermore, while computing the closest feasible action results in a high sample efficiency (as in \texttt{OptLayer}), it does not necessarily have a high utility (especially in the initial learning stage of RL agents). In contrast, providing a safe fallback policy \textit{a priori} can lead to a high initial utility, but was shown to result in a poor sample efficiency and inability to include equality constraints (as in \texttt{SafeFallback}). In this paper, we present two novel advancements: (I) combining the \texttt{OptLayer} and \texttt{SafeFallback} method, named \texttt{OptLayerPolicy}, to increase the initial utility while keeping a high sample efficiency and the possibility to formulate equality constraints. (II) introducing self-improving hard constraints, to increase the accuracy of the constraint functions as more and new data becomes available so that better policies can be learnt. Both advancements keep the constraint formulation decoupled from the RL formulation, so new (presumably better) RL algorithms can act as drop-in replacements. We have shown that, in a simulated multi-energy system case study, the initial utility is increased to 92.4\% (\texttt{OptLayerPolicy}) compared to 86.1\% (\texttt{OptLayer}) and that the policy after training is increased to 104.9\% (\texttt{GreyOptLayerPolicy}) compared to 103.4\% (\texttt{OptLayer}) - all relative to a \textit{vanilla} RL benchmark. Although introducing surrogate functions into the optimisation problem requires special attention, we conclude that the newly presented \texttt{GreyOptLayerPolicy} method is the most advantageous.}


\usepackage{graphicx}
\usepackage{pdflscape}
\usepackage{longtable}
\usepackage{cleveref}
\usepackage{wrapfig}
\usepackage[ruled,vlined]{algorithm2e}
\usepackage[official]{eurosym}
\usepackage{makecell}
\usepackage{adjustbox}
\usepackage{lineno}
\usepackage{float}
\floatstyle{plaintop}
\restylefloat{table}
\usepackage{nomencl}
\usepackage{etoolbox}
\usepackage{multicol}
\makenomenclature

\renewcommand\nomgroup[1]{%
  \item[\bfseries
  \ifstrequal{#1}{Y}{Subscripts/Superscripts}{%
  \ifstrequal{#1}{B}{Symbols}{%
  \ifstrequal{#1}{A}{Sets}{%
  \ifstrequal{#1}{Z}{Other symbols}{}}}}%
]}

\DeclareMathOperator*{\argmin}{arg\,min}
\begin{document}
%%%%%%%%%%%%%%%%%%%%%%%%%%%%%%%%%%%%%%%%%%

\section*{Highlights} %% 3 to 5 bullet points (maximum 85 characters, including spaces, per bullet point).
\begin{itemize}
    \item Increased initial utility while retaining a high sample efficiency
    \item Ability to self-improve the hard constraints 
    \item Better policies can be found with more accurate constraints
    \item Constraint formulation remains decoupled from (optimal) control
\end{itemize}

%\newpage
%%%%%%%%%%%%%%%%%%%%%%%%%%%%%%%%%%%%%%%%%%
%\setcounter{section}{-1} %% Remove this when starting to work on the template.

\nomenclature[B]{\(x, s\)}{State} 
\nomenclature[A]{\(\mathbb{X}, S\)}{State space}
\nomenclature[A]{\(\mathbb{X}_c\)}{State constraints}
\nomenclature[B]{\(u, a\)}{Action}
\nomenclature[A]{\(\mathbb{U}, A\)}{Action space}
\nomenclature[A]{\(\mathbb{U}_c\)}{Action constraints}
\nomenclature[B]{\(w\)}{Wiener process}
\nomenclature[B]{\(\pi\)}{Policy}
\nomenclature[B]{\(\gamma\)}{Discount factor, Binary variable}
\nomenclature[B]{\(C\)}{Cost}
\nomenclature[B]{\(z\)}{Infeasibility cost}
\nomenclature[B]{\(c\)}{Constraint}
\nomenclature[B]{\(E\{ \cdot \}\)}{Expectation}
\nomenclature[B]{\(L\)}{Loss}
\nomenclature[B]{\(t\)}{Time}
\nomenclature[A]{\(\mathbb{T}\)}{Time domain}
\nomenclature[B]{\(J\)}{Objective}
\nomenclature[B]{\(R\)}{Reward}
\nomenclature[A]{\(\mathbb{R}\)}{Real numbers}
\nomenclature[B]{\(n\)}{Amount}
\nomenclature[B]{\(h\)}{Threshold}
\nomenclature[B]{\(d\)}{Distance}
\nomenclature[Y]{\(i\)}{initial}
\nomenclature[Y]{\(j\)}{constraint index}
\nomenclature[Y]{\(safe\)}{safe, feasible}
\nomenclature[Y]{\('\)}{next}
\nomenclature[Z]{\(\tilde{(\cdot)}\)}{Predicted}
\nomenclature[Z]{\(\Bar{(\cdot)}\)}{Nominal, Average}
\nomenclature[Z]{\(\Hat{(\cdot)}\)}{Unknown}
\nomenclature[B]{\(Q\)}{Thermal power}
\nomenclature[B]{\(P\)}{Electrical power}
\nomenclature[B]{\(\eta\)}{Efficiency}
\nomenclature[B]{\(COP\)}{Coefficient of performance}
\nomenclature[B]{\(SOC\)}{State of charge}
\nomenclature[B]{\(T\)}{Temperature}
\nomenclature[Y]{\(min\)}{Minimum}
\nomenclature[Y]{\(max\)}{Maximum}
\nomenclature[Y]{\(boil\)}{Boiler}
\nomenclature[Y]{\(hp\)}{Heat pump}
\nomenclature[Y]{\(chp\)}{Combined heat and power}
\nomenclature[Y]{\(tess\)}{Thermal energy storage system}
\nomenclature[Y]{\(bess\)}{Battery energy storage system}
\nomenclature[Y]{\(evap\)}{Evaporator}
\nomenclature[Y]{\(cond\)}{Condenser}
\nomenclature[Y]{\(env\)}{Outdoor environment}
\nomenclature[Y]{\(th\)}{Thermal}
\nomenclature[Y]{\(el\)}{Electrical}
\nomenclature[Y]{\(wind\)}{Wind turbines}
\nomenclature[Y]{\(solar\)}{Photovoltaic system}
\nomenclature[Y]{\(train\)}{Training (interval)}
\nomenclature[B]{\(X\)}{Price}
\nomenclature[B]{\(H\)}{Hour of the day}
\nomenclature[B]{\(D\)}{Day of the week}
\nomenclature[B]{\(Pr\)}{Probability}
\nomenclature[B]{\(P_a\)}{Transition probability}
\nomenclature[B]{\(x, y\)}{Scalarisation weights}

\begin{multicols}{2}
\printnomenclature
\end{multicols}

\section{Introduction}
\par The possibility to integrate multiple energy, commodity, and utility streams is increasingly available as the energy technologies that allow for this sector coupling are more mature, more widely spread installed, and more creatively found. The overall system efficiency and performance can then be enhanced by implementing an integrated control strategy, that considers all energy assets across all energy carriers – including all sources of flexibility (i.e., storage, controllable loads) within all sub-systems. This seemingly limitless optimisation potential then typically has an economic or environmental-oriented objective \cite{Fabrizio2009Trade-offSystems} or has a combination of multiple, sometimes conflicting, objectives.

\par Finding an optimum or Pareto optimum level of operation for such multi-energy systems is no small task. It requires establishing and maintaining specific set-points to first ensure a disruptive-free operation by fulfilling all system constraints and secondary to pursue a desired objective (e.g., energy costs or $CO_{2}$-equivalent emissions minimisation) \cite{Engell2007FeedbackOperation}. Moreover, the use of flexibilities introduces a dependency between successive time steps, which in theory requires an \textit{infinite} horizon optimisation calculation. While managing multiple uncertainties (e.g., variation in demands, weather, and pricing) only leaves us with an \textit{expectation} in these \textit{continuous} systems.

\par In practice, model-predictive control (MPC) is often used as the optimal control technique as it has mature stability, feasibility, robustness, and constraint handling theory \cite{Gorges2017RelationsLearning}. However, it does require a detailed model in advance (i.e., plant models, input and output disturbance models, measurement noise models and prediction models for states not included in the plant model - e.g., demand forecasts, weather forecasts, price forecasts) which is typically not adaptive \cite{Ceusters2021Model-predictiveStudies}. In an attempt to overcome these shortcomings, reinforcement learning (RL) is model-free and inherently adaptive, yet has immature stability, feasibility, robustness, and constraint handling theory. It is only recently that \citeauthor{Ceusters2023SafeFunctions} \cite{Ceusters2023SafeFunctions} showed that a (near-to) optimal multi-energy management policy can be learnt safely. Hereby, the project-specific upfront and ongoing engineering efforts remain reduced, a better representation of the underlying system dynamics can still be learnt, and modelling bias is kept to a minimum (no model-based objective function). 

\subsection{Problem statement}

\par
\begin{figure}[H]
    \centering
    \begin{subfigure}{.5\textwidth}
        \centering
        \includegraphics[width=1\linewidth]{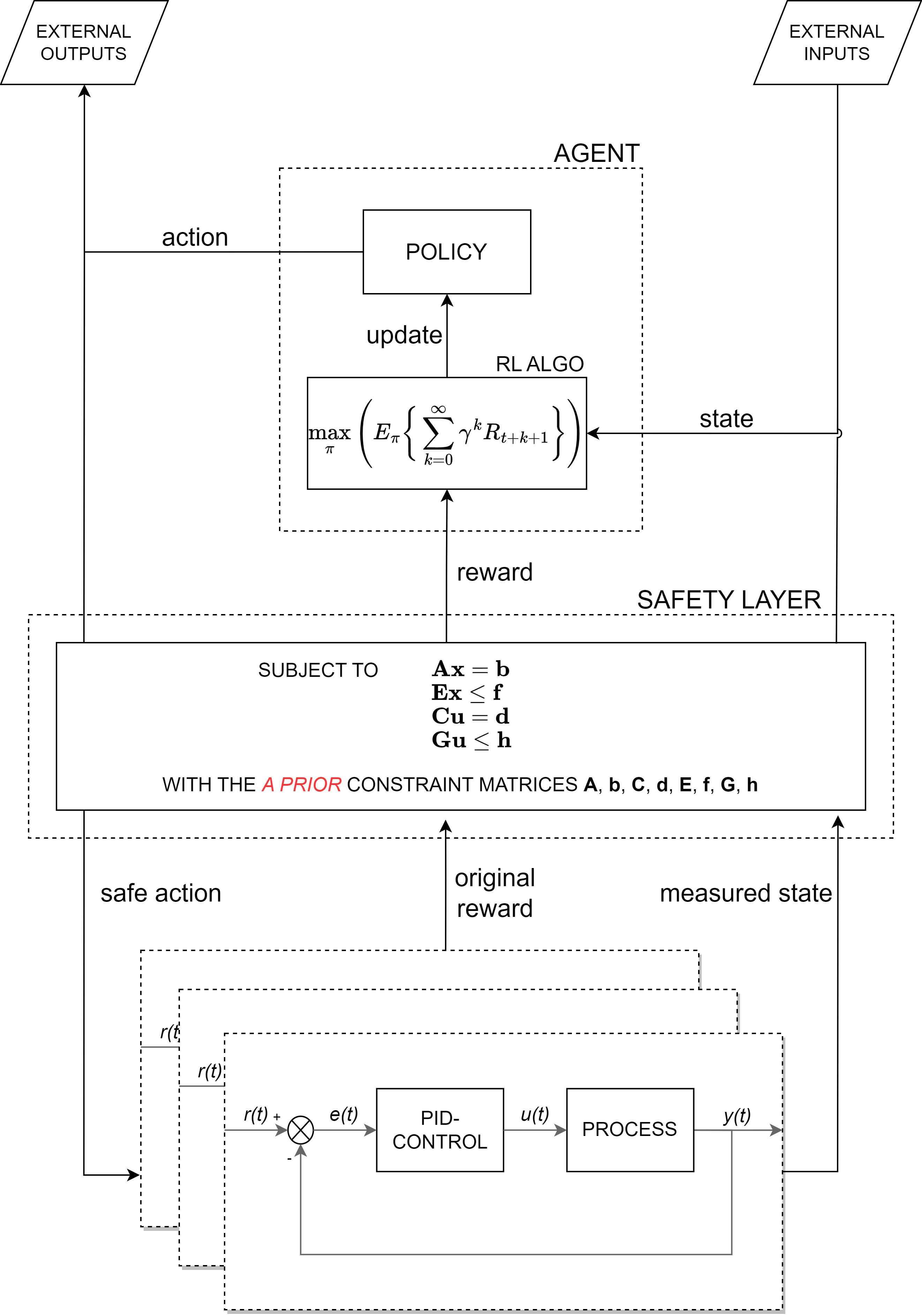}
        \caption{\textit{vanilla} shielded (safe) reinforcement learning}
        \label{fig: vanillasaferl}
    \end{subfigure}%
    \begin{subfigure}{.5\textwidth}
        \centering
        \includegraphics[width=1\linewidth]{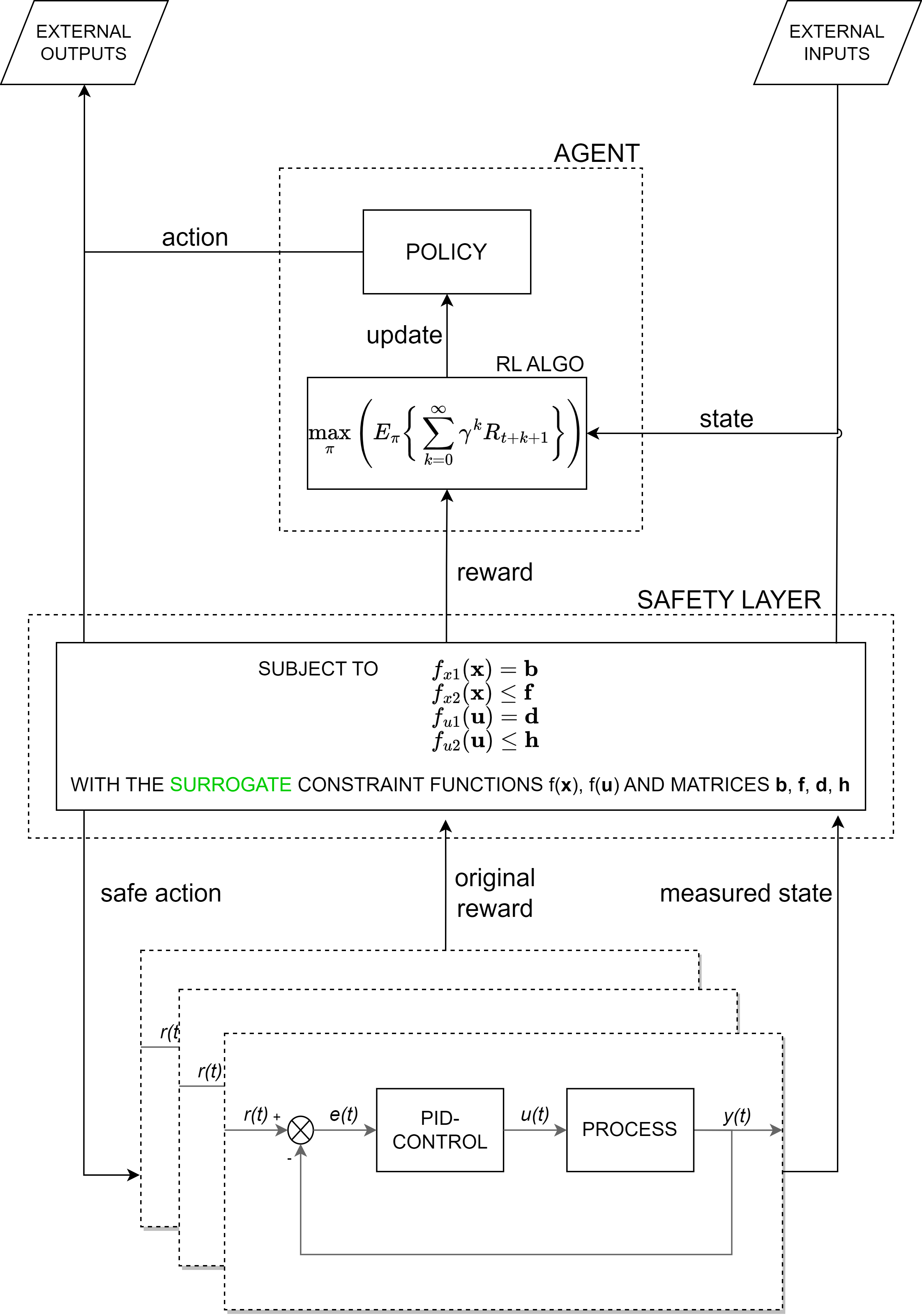}
        \caption{\textit{grey} shielded (safe) reinforcement learning}
        \label{fig: greysaferl}
    \end{subfigure}
    \vspace*{10pt}
    \caption{Block diagram representation: the actions of the RL agent, when in a specific state, are checked for feasibility by evaluating them against a set of, either \textit{a priori} or \textit{surrogate}, constraint functions. These functions then serve as a safety layer measure to protect the environment from unsafe control actions. It is important to note that we assume that the continuous handling of unconstrained errors (which involves minimising the difference between a desired set-point and a measured process variable) is carried out by proportional-integral-derivative (PID) controllers.}
    \label{fig:blockdiagrams}
\end{figure}

 However, even the constraint functions themselves (see \autoref{fig: vanillasaferl}) are not always trivial to provide accurately in advance (e.g., see the energy balance constraint \autoref{equation9d}) – especially when auxiliary state variables are required or when they need to be activated under particular conditions \cite{Pham2018OptLayerWorld}. Even when they could be given in advance, they may not continue to be sufficiently accurate over time (e.g., efficiency decay). Furthermore, while computing the closest feasible action results in a high sample efficiency (as in \texttt{OptLayer} \cite{Pham2018OptLayerWorld}), it does not necessarily have a high utility (especially in the initial learning stage of RL agents). In contrast, providing a safe fallback policy \textit{a priori} can lead to a high initial utility, but was shown to result in a poor sample efficiency and the inability to include equality constraints (as in \texttt{SafeFallback} \cite{Ceusters2023SafeFunctions}).

\par Our goal, therefore, is not only to ensure that \textit{every} interaction with the underlying environment (a multi-energy system in our case study) is safe by satisfying to a set of constraints. But also to improve the accuracy of the constraints themselves (see \autoref{fig: greysaferl}), as more data becomes available, to improve the initial utility and keep a high sample efficiency while the constraint handling remains independent of the (optimal) control technique. This is so that future, presumably better, yet inherently unsafe, optimisation algorithms (e.g., a new RL algorithm) can act as drop-in replacements.

\subsection{Contribution and outline}
\par Our contributions presented in this work are, to the best of the authors' knowledge, believed to be the first of their kind, and can be outlined as follows:

\begin{itemize}
    \item Combining the \texttt{OptLayer} \cite{Pham2018OptLayerWorld} and \texttt{SafeFallback} \cite{Ceusters2023SafeFunctions} method, named \texttt{OptLayerPolicy}, to increase the initial utility while keeping a high sample efficiency and the possibility to formulate equality constraints.
    \item Introducing self-improving (adaptive) hard constraints, to increase the accuracy of the constraint functions as more and new data becomes available so that better policies can be learnt.
\end{itemize}

\par In \autoref{chapter2} we have a concise discussion of related work, \autoref{chapter3} introduces the proposed methodologies, while \autoref{chapter4} presents the case study specific toolchain, multi-energy system simulation environment, safety layer, RL agent, and evaluation procedure. Finally, \autoref{chapter5} discusses the results and provides directions for future work so that \autoref{chapter6} presents our conclusion. In \hyperref[Appendix A]{Appendix~\ref*{Appendix A}}  we show time series visualisations of the self-discovered policies, in \hyperref[Appendix B]{Appendix~\ref*{Appendix B}} the full learning and cost curves for the agents assessed in the case study, in \hyperref[Appendix C]{Appendix~\ref*{Appendix C}} the pseudocode and hyperparameters of the specific RL agent (TD3, i.e., Twin Delayed Deep Deterministic Policy Gradient) and in \hyperref[Appendix D]{Appendix~\ref*{Appendix D}} the run-time statistics.

\section{Related work} \label{chapter2}
\par Reinforcement learning has been proposed and demonstrated for a wide variety of applications in power and energy systems, as extensively reviewed by e.g. \citeauthor{Cao2020ReinforcementReview} \cite{Cao2020ReinforcementReview}, \citeauthor{Yang2020ReinforcementSurvey} \cite{Yang2020ReinforcementSurvey} and \citeauthor{Perera2021ApplicationsSystems} \cite{Perera2021ApplicationsSystems}, and even for multi-energy systems more specifically by \citeauthor{Zhou2022AdvancesPerspectives} \cite{Zhou2022AdvancesPerspectives}. It ranges from real-time control (e.g., robust voltage control by \citeauthor{Petrusev2023ReinforcementUncertainties} \cite{Petrusev2023ReinforcementUncertainties}) to energy management systems (EMS) RL applications. Recent EMS examples include, \citeauthor{Zhou2022Data-drivenLearning} \cite{Zhou2022Data-drivenLearning} who proposed deep RL for the stochastic EMS of a multi-energy system and introduced a prioritised experience relay that improves the training efficiency and thus the convergence rate of the RL algorithm. While a multi-agent deep RL EMS, using multi-agent counterfactual soft actor-critic (mCSAC \cite{Pu2021DecomposedLearning}), was demonstrated by \citeauthor{Zhu2022EnergyPark} \cite{Zhu2022EnergyPark} in a simulated multi-energy industrial park, by \citeauthor{Ahrarinouri2022DistributedHubs} \cite{Ahrarinouri2022DistributedHubs} using multi-agent Q-learning in a simulated distributed and interconnected multi-carrier energy hub case study, and by \citeauthor{Jendoubi2022Data-drivenLearning} \cite{Jendoubi2022Data-drivenLearning} using multi-agent Deep Deterministic Policy Gradient (MADDPG) in separate simulated microgrid, eco-neighbourhood and flat building case studies. \citeauthor{Ceusters2021Model-predictiveStudies} \cite{Ceusters2021Model-predictiveStudies}, furthermore, benchmarked an on- and off-policy multi-objective model-free deep RL algorithm against a linear MPC on two separate simulated multi-energy systems and showed that the RL agent, using soft constraints, can outperform the MPC (since it learnt a better representation of the \textit{true} system dynamics). \citeauthor{Sun2023Multi-objectiveAlgorithm} \cite{Sun2023Multi-objectiveAlgorithm} then also demonstrated a multi-objective deep RL approach with soft constraints yet on an IEEE-30 node optimal power flow problem. However, using RL, without adequate measures, could result in undesirable specific losses (e.g., monetary, comfort) and, in extreme cases, in human harm. This as, RL inherently requires the interaction with its environment - and does this without the consideration of any constraints (besides the limits of the action space itself). As also reported by \citeauthor{Ceusters2023SafeFunctions} \cite{Ceusters2023SafeFunctions}, most works knowingly (and therefore reported as such) or unknowingly, either neglected these environment-specific constraints or greatly simplified them - limiting their real-world use. Safe RL, therefore, aims to: "\textit{learn policies that maximise the expectation of the return in problems in which it is important to ensure reasonable system performance and/or respect safety constraints during the learning and/or deployment processes}" - as defined by \citeauthor{Garcia2015ALearning} \cite{Garcia2015ALearning}. Then \citeauthor{Ceusters2023SafeFunctions} \cite{Ceusters2023SafeFunctions} were one of the first to show that a (near-to) optimal multi-energy management policy can be learnt safely with hard constraints, that these constraints can be formulated independently of the (optimal) control technique, and that better policies can be found starting from an initial safe fallback policy. While \citeauthor{Feng2023EconomicApproach} \cite{Feng2023EconomicApproach} introduced a robust state generation procedure in combination with a dynamic pricing mechanism to economically dispatch an industrial park using a variation of the distributed proximal policy optimisation (DPPO) algorithm subject to a set of hard constraints to ensure safe (near-to) optimal operation. However, even only assuming the availability of perfectly accurate constraint functions themselves is not always possible – especially when auxiliary state variables are required or when they need to be activated under particular conditions, as also needed by \texttt{OptLayer} \cite{Pham2018OptLayerWorld} which is considered a current state-of-the-art benchmark.

\par On the other hand, more traditional optimisation approaches - typically in a receding horizon control manner (e.g., model-predictive control), still see significant advancements for the management of multi-energy systems. Recent examples include: \citeauthor{Zhu2020EnergyNetworks} \cite{Zhu2020EnergyNetworks} who designed a joint multi-energy scheduling and trading algorithm based on Lyapunov optimisation and a double-auction mechanism. Simulations based on real data showed that individual microgrids could achieve a time-averaged profit that was arbitrarily close to an optimum value while avoiding compromising their own comfort. \citeauthor{Zhu2022StochasticConstraints} \cite{Zhu2022StochasticConstraints} then later employed a fast distributed algorithm based on stochastic gradient descent with a two-timescale implementation to address energy storage constraints and short-term balancing. In addition, they estimated users' willingness to shift their load, who participated in an incentive mechanism to reduce peak loads. Analytical and numerical results showed that when the bid-ask spread of electricity was sufficiently small, the proposed algorithm could achieve cost levels close to optimal asymptotically. Also, \citeauthor{Zou2023Peer-to-PeerNetwork} \cite{Zou2023Peer-to-PeerNetwork} introduced a two-tiered peer-to-peer (P2P) multi-energy trading system for an interconnected distribution network (DN) and district heating network (DHN). In the lower tier, nodal agents optimised energy schedules and P2P trading strategies using Nash bargaining theory. In the upper tier, operators minimised power losses and ensured network constraints by reconfiguring the DN and DHN and adjusting trades as needed. The DN operation was modelled using a linearised DistFlow model with radiality constraints, while the DHN employed a quasi-linear thermal flow model. The framework's effectiveness was demonstrated on an IEEE 33-bus DN and a 23-node DHN. And as a final example, \citeauthor{Zou2023AMicrogrid} \cite{Zou2023AMicrogrid} introduced an energy management method for a multi-energy microgrid (MEMG), that employed the transactive energy concept and formulated the problem as a Stackelberg game-theoretic bi-level optimisation model. This approach formed a day-ahead stochastic mixed-integer linear program (MILP) and an intra-day deterministic model. To solve these models, an adaptive Progressive Hedging algorithm decomposed the day-ahead stochastic MILP into scenario-based subproblems that could be solved in parallel. Meanwhile, an outer approximation algorithm was employed in the intra-day stage to linearise the bi-linear objective function. Nevertheless, it is clear that these approaches require highly detailed mathematical formulations \textit{a priori} (i.e., plant models, input and output disturbance models, measurement noise models and prediction models for states not included in the plant model - e.g., demand forecasts, weather forecasts, price forecasts), which are typically not adaptive and which are not all required with safe RL \cite{Ceusters2021Model-predictiveStudies}.

\par Considering a broader view across both the RL research space and the control theory space, \citeauthor{Brunke2021SafeLearning} \cite{Brunke2021SafeLearning} provided a safe learning review and showed: (1) approaches that learn uncertain system dynamics and safely improve the policy starting with an imperfect \textit{a priori} dynamic model, (2) approaches that do not have a model or even constraints in advance and encourage safety or robustness (e.g., by penalising dangerous actions) but provide no strict guarantees, and (3) approaches that provide safety certificates to inherently unsafe learning-based controllers, using an \textit{a priori} dynamic model. Hence, there are multiple approaches to safe (reinforcement) learning that exist, varying in their level of safety. These approaches can be categorised as follows, in ascending order of safety: soft-constraint satisfaction, chance-constraint satisfaction, and hard-constraint satisfaction. Recent examples – one of each category – include, \citeauthor{McKinnon2020Context-awareControl} \cite{McKinnon2020Context-awareControl} who proposed a stochastic MPC, where the predicted cost, using a computationally efficient yet expressively limited \textit{ a priori} dynamic model, is corrected by a simple learnt dynamics model over the MPC horizon. \citeauthor{Bharadhwaj2021ConservativeExploration} \cite{Bharadhwaj2021ConservativeExploration} who extended Conservative Q-Learning (CQL) towards Conservative Safety Critic (CSC) and showed safety constraint satisfaction with \textit{high probability} while providing provable safe policy improvements. And finally, \citeauthor{Lopez2021RobustSafety} \cite{Lopez2021RobustSafety} who introduced robust adaptive control barrier functions (CBF) which allowed safe adaptation of structured parametric uncertainties in the time derivatives of CBFs which are used together with an inherently unsafe adaptive control algorithm.

\par Nevertheless, a model-free safe RL approach of the following \textit{combined} characteristics has – to the best of the authors' knowledge – never been proposed: (i) providing hard-constraints satisfaction guarantees (ii) while decoupled from the RL (as a Markov Decision Process) formulation, (iii) both during training a (near) optimal policy (involving exploratory and exploitative steps) as well as during the deployment of any policy (e.g., offline pre-trained RL agents) and (iv) this while learning uncertain constraint components and safely improving the policy with high sample efficiency, and (v) starting from an increased initial utility, to (vi) demonstrate for the energy management of multi-energy systems.

\section{Proposed methodology} \label{chapter3}
\par We start from a discrete time-varying stochastic system and recognise that this is an approximation for continuous\footnote{as we assume that the continuous error handling is performed by well-tuned PID-controllers} systems, in the form of:

\begin{equation}
    x_{t+1}=f_t(x_t,\ u_t,\ w_t)
    \label{equation1}
\end{equation}
where \(x_t\) is the \textit{n}-dimensional \textit{state} vector, which is an element of the \textit{state space} \(\mathbb{X}\), and \(u_t\) the \textit{m}-dimensional \textit{control} or \textit{action} vector, which is an element of the input or \textit{action space} \(\mathbb{U}\), and \(w_t\) a Wiener process, i.e. some stochastic noise, that – with this formulation – can enter the dynamics in any form. The problem is to find a control signal \( u_t = \pi_t(x_t) \), with \( \pi \) being the control \textit{policy}, so that the infinite-horizon, yet discounted (if \( \gamma^t < 1\)), cost function from \citeauthor{Gorges2017RelationsLearning} \cite{Gorges2017RelationsLearning} in a unified notation

\begin{align}
    C( x_i,\ u_t) &=  E \bigg\{\sum_{t = 0}^{\infty} \gamma^t L_t(x_t,u_t,w_t) \bigg\}
    \label{equation2}
\end{align}
is minimal, where  \( L_t(x_t,u_t,w_t) \) is the stage \textit{loss}. Considering the inherently stochastic nature of the system described by \autoref{equation1}, one can only strive to minimise the \textit{expectation} \( E\{ \cdot \} \) of the corresponding cost function across the multitude of stochastic trajectories originating from \(x_i\).
This results in the discounted infinite-horizon objective function, again from \citeauthor{Gorges2017RelationsLearning} \cite{Gorges2017RelationsLearning} in unified notation:

\begin{equation}
    J_\infty(x_t)= \min_{u_t \ \in \ U_t} \Bigg( E \bigg\{\sum_{t = 0}^{\infty} \gamma^t L_t(x_t,u_t, w_t) \bigg\} \Bigg)\ , \ \forall \ x_t 
    \label{equation3}
\end{equation}

\par Following the standard RL formulation of the state-value function from \citeauthor{Sutton2018ReinforcementIntroduction} \cite{Sutton2018ReinforcementIntroduction}, the objective is to find a policy \( \pi \), which is a mapping of states, \( s = x \), to actions, \(a = u\), that maximises an expected sum of discounted rewards, yet making it subject to the constraints, \( c \):

\begin{subequations}
\begin{align}
    \label{equation4a}
    \max_\pi &\Bigg(E_\pi\bigg\{ \sum_{k = 0}^{\infty} \gamma^k R_{t+k+1} \bigg\}\Bigg)\\[1em]
    \label{equation4b}
    s.t. \quad & s_t = s && t \in \mathbb{T}_{0}^{+\infty} \\
    \label{equation4c}
    \quad & c_t^j(s_t, a_t, w_t) \leq 0 && t \in \mathbb{T}_{0}^{+\infty}, j \in \mathbb{R}_{1}^{n_c}
\end{align}
\end{subequations}
where \(E_\pi\) is the expected value, following the policy \(\pi\), of the rewards \(R\) and reduced with the discount factor \(\gamma\) over an infinite sum at any time step \(t\) with \(n_c\) amount of constraint functions.

\par Note that \autoref{equation4a} corresponds to a stochastic optimal control problem with an infinite time horizon in discrete time. It can be regarded as the time-invariant version of \autoref{equation3}, where \( R_t = -L_t(x_t,u_t, w_t) \). However, it diverges from the standard formulation of RL due to the subjection of hard-constraint functions, \(c_t^j\).
This can include \textit{state constraints} \( \mathbb{X}_c \in \mathbb{X} \), action or \textit{input constraints} \( \mathbb{U}_c \in \mathbb{U} \) and \textit{stability guarantees} (e.g., Lyapunov, asymptotic or exponential stability). Rather than proposing a specific safe RL algorithm, we propose to decouple the constraint function formulation from the (RL) agent so that any (new RL) algorithm can be used – while always guaranteeing hard-constraint satisfaction and this in a minimally invasive way (i.e., correcting actions to the closest possible feasible action).

\subsection{OptLayerPolicy method} \label{chapter 3.1}
\par The minimal invasive correction of predicted actions, \(\tilde a\), e.g. originating from a RL algorithm, can be expressed from \citeauthor{Pham2018OptLayerWorld} \cite{Pham2018OptLayerWorld} (i.e., the original \texttt{OptLayer} formulation) as:

\begin{subequations}
\begin{align}
    \label{equation5a}
    a_{safe, t} = &\argmin_{a_t} \frac{1}{2} \parallel a_t - \tilde a_t \parallel^2 \\[1em]
    \label{equation5b}
    s.t. \quad & c_t^j(s_t, a_t, w_t) \leq 0 && t \in \mathbb{T}_{0}^{+\infty}, j \in \mathbb{R}_{1}^{n_c}
\end{align}
\end{subequations}
which is a Quadratic Program (QP), where \( a_t \) is the optimisation variable that results in the closest feasible safe action \( a_{safe} \), which does not affect the optimality as \(a_t\) does not depend on \(\tilde a_t\) \cite{Pham2018OptLayerWorld}. The distance closest to the possible feasible action is then simply:

\begin{equation}
    \label{equation6}
    d_{safe, t} = \min_{a_t \in \mathbb{R}^{n_c}} \frac{1}{2} \parallel a_t - \tilde a_t \parallel^2 
\end{equation}
While this is the closest distance for a minimally invasive correction, this does not necessarily result in a close-to-optimal action. We, therefore, propose to fallback on an \textit{a priori} safe policy, \( \pi^{safe} \) (i.e., as in the \texttt{SafeFallback} \cite{Ceusters2023SafeFunctions} algorithm), when the distance, \( d_{safe} \), surpasses a given threshold \( h_{safe} \). This safe fallback policy can typically be derived through classic control theory in the form of a set of hard-coded rules such as a simple rule-based policy (e.g., a priority-based energy management strategy - which is commonly available or easily constructable - see \autoref{chapter4.4} for the safe fallback policy of the considered case study).

While this relies on the utility of the (non-optimal) safe fallback policy itself, we will later show its effectiveness. The \texttt{OptLayerPolicy} algorithm is shown in \autoref{algo: optlayerpolicy}.

\begin{algorithm}
\DontPrintSemicolon
\SetAlgoLined
 \nl Input: initialise RL algorithm, initialise constraint functions in sets \( \mathbb{X}_c \) and \( \mathbb{U}_c \), initialise safe fallback policy \(\pi^{safe}\) \;
 \nl \For{\( k=0,1,2,\dots \)}{
 \nl Observe state \(s\) and predict action \( \tilde a\)\;
 \nl Compute safety distance \(d_{safe}\)\;
 \nl \eIf{\(d_{safe} \leq h_{safe}\)}{\( a_{safe} = \argmin\limits_{a_t} \frac{1}{2} \parallel a_t - \tilde a_t \parallel^2 \quad s.t. \quad c_t^j(s_t, a_t, w_t) \leq 0 \)}{\( a_{safe} = \pi^{safe}(s)\)}
 \nl Execute \( a_{safe} \) in the environment \;
 \nl Observe next state \( s' \), reward \( r \) and done signal \( d \) to indicate whether \( s' \) is terminal \;
 \nl Give experience tuple \( (s,a_{safe},r,s',d) \) \textbf{and if} \(a_{safe} \neq \tilde a:\) \( (s,\tilde a,r-z,s',d) \) with cost \(z\) \;
 \nl If \( s' \) is terminal, reset environment state \;
 }
 \caption{OptLayerPolicy}\label{algo: optlayerpolicy}
\end{algorithm}

\subsection{GreyOptLayerPolicy method} \label{chapter 3.2}
\par We can improve the initial utility of the \textit{vanilla} RL agent, by using a safe fallback policy when a predicted unsafe action is too far from the feasible solution space (i.e., under the notion that random\footnote{as RL agents typically have an initial random exploration phase} safe actions have a low utility) - as outlined in the previous section. However, all constraint functions (\autoref{equation5b}) are assumed to be \textit{true}. In reality, we typically only have access to a nominal set of constraints with an \textit{a priori} unknown error (e.g., see \autoref{equation9d} and \autoref{tab: safety layer model metrics}). We can express this from \citeauthor{Brunke2021SafeLearning} \cite{Brunke2021SafeLearning} for constraint functions as:

\begin{equation}
    \label{equation7}
    c_t^j(s_t, a_t, w_t) = \Bar{c}_t^j(s_t, a_t) + \Hat{c}_t^j(s_t, a_t, w_t)
\end{equation}
where \( \Bar{(\cdot)} \) is the nominal component, reflecting our prior knowledge, and \( \Hat{(\cdot)} \) is an unknown component, that can be learnt from data - making the equation and therefore the approach \textit{adaptive}. The \texttt{GreyOptLayerPolicy} algorithm then becomes:

\begin{algorithm}
\DontPrintSemicolon
\SetAlgoLined
 \nl Input: initialise RL algorithm, initialise nominal constraint functions \( \Bar{c}_t^j \) in sets \( \mathbb{X}_c \) and \( \mathbb{U}_c \), initialise supervised learning models for \( \Hat{c}_t^j \), initialise safe fallback policy \(\pi^{safe}\) \;
 \nl \For{\( k=0,1,2,\dots \)}{
 \nl Observe state \(s\) and predict action \( \tilde a\)\;
 \nl Compute safety distance \(d_{safe}\)\;
 \nl \eIf{\(d_{safe} \leq h_{safe}\)}{\( a_{safe} = \argmin\limits_{a_t} \frac{1}{2} \parallel a_t - \tilde a_t \parallel^2 \quad s.t. \quad \Bar{c}_t^j(s_t, a_t) + \Hat{c}_t^j(s_t, a_t, w_t) \leq 0 \)}{\( a_{safe} = \pi^{safe}(s)\)}
 \nl Execute \( a_{safe} \) in the environment \;
 \nl Observe next state \( s' \), reward \( r \) and done signal \( d \) to indicate whether \( s' \) is terminal \;
 \nl Give experience tuple \( (s,a_{safe},r,s',d) \) \textbf{and if} \(a_{safe} \neq \tilde a:\) \( (s,\tilde a,r-z,s',d) \) with cost \(z\) \;
 \nl \If{\(k \mod h_{train} = h_{train} - 1 \)}{fit \( \Hat{c}_t^j \) using supervised learning with buffer of \( (s, a_{safe}) \)}
 \nl \textbf{if} \( s' \) is terminal, reset environment state \;
 }
 \caption{GreyOptLayerPolicy}\label{algo: greyoptlayerpolicy}
\end{algorithm}

where \(h_{train}\) is the training interval for fitting the \textit{a priori} unknown components of the constraints (see \autoref{chapter4.4} for the specifics of our case study). However, this introduces function approximators (e.g., artificial neural networks) into the optimisation problem (\autoref{equation5a}) – which are not trivial to integrate when using exact solving methods. Nevertheless, \citeauthor{Gunnell2022MachineConstraints} \cite{Gunnell2022MachineConstraints} recently integrated multiple machine learning algorithms in a gradient descent optimisation framework - which is also the framework used in this work (i.e., \texttt{GEKKO} \cite{Beal2018GEKKOSuite}).

\section{Case study} \label{chapter4}
\subsection{Toolchain}
\par
\begin{figure}[H]
    \centering
    \includegraphics[scale=0.125]{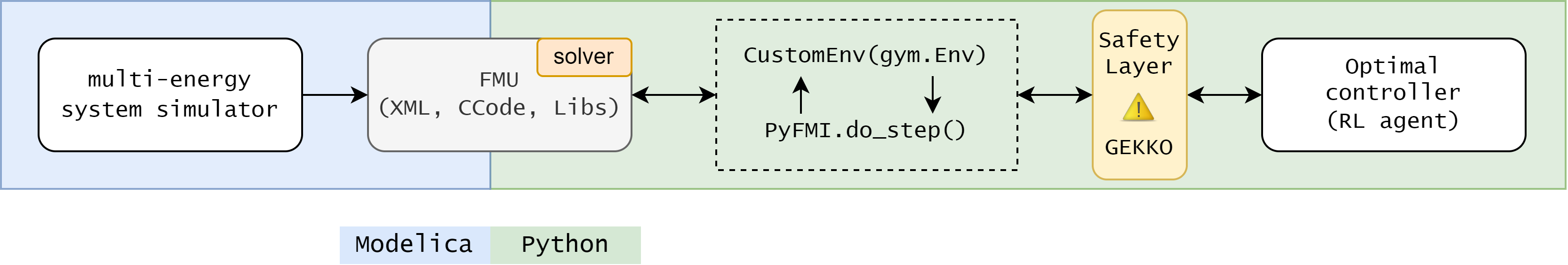}
    \caption{Toolchain architecture. Note that we used the \texttt{do\_step()} method in \texttt{PyFMI} \cite{Andersson2016PyFMI:Interface} over \texttt{simulate()} due to the significant reduction in run-time when initialised correctly, and also note that the Differential Algebraic Equations solver is part of the co-simulation FMU.}
    \label{fig: architecture}
\end{figure}

 We use a multi-energy systems simulation model, that first was developed by \citeauthor{Ceusters2021Model-predictiveStudies} \cite{Ceusters2021Model-predictiveStudies} and later modified by \citeauthor{Ceusters2023SafeFunctions} \cite{Ceusters2023SafeFunctions}. This allowed for the consequence-free verification of the safe operation (i.e., with no risk of violating real-life constraints with its potential loss of comfort or, in extreme cases, human harm). The presumed to be \textit{true} multi-physical first-principle equations were developed in \texttt{Modelica} \cite{Mattsson1998PhysicalModelica} due to its object-oriented nature and the availability of highly specialised libraries and elementary components. To allow for the exchange across different simulation environments and programming languages, this dynamic system model was exported as a co-simulation \textit{functional mock-up unit} (FMU), as also proposed by \citeauthor{Graber2017FromProblems} \cite{Graber2017FromProblems} and then wrapped in an \texttt{OpenAI} gym \cite{Brockman2016OpenAIGym} in \texttt{Python}. The mixed-integer quadratic problem in the safety layer was formulated with \texttt{GEKKO} \cite{Beal2018GEKKOSuite}, as it allowed for the integration of machine learning algorithms (used for \( \Hat{(\cdot)} \) in \autoref{equation7}) into an exact optimisation framework. We have used \texttt{Scikit-learn} \cite{PedregosaFABIANPEDREGOSA2011Scikit-learn:Python} for the supervised learning models of \( \Hat{c}_t^j \). The toolchain architecture is shown in \autoref{fig: architecture}.

\subsection{Simulation model} \label{chapter4.2}
\par The simulated multi-energy system, from \citeauthor{Ceusters2023SafeFunctions} \cite{Ceusters2023SafeFunctions}, has the following topology:

\begin{figure}[H]
	\centering
	\includegraphics[scale=0.09]{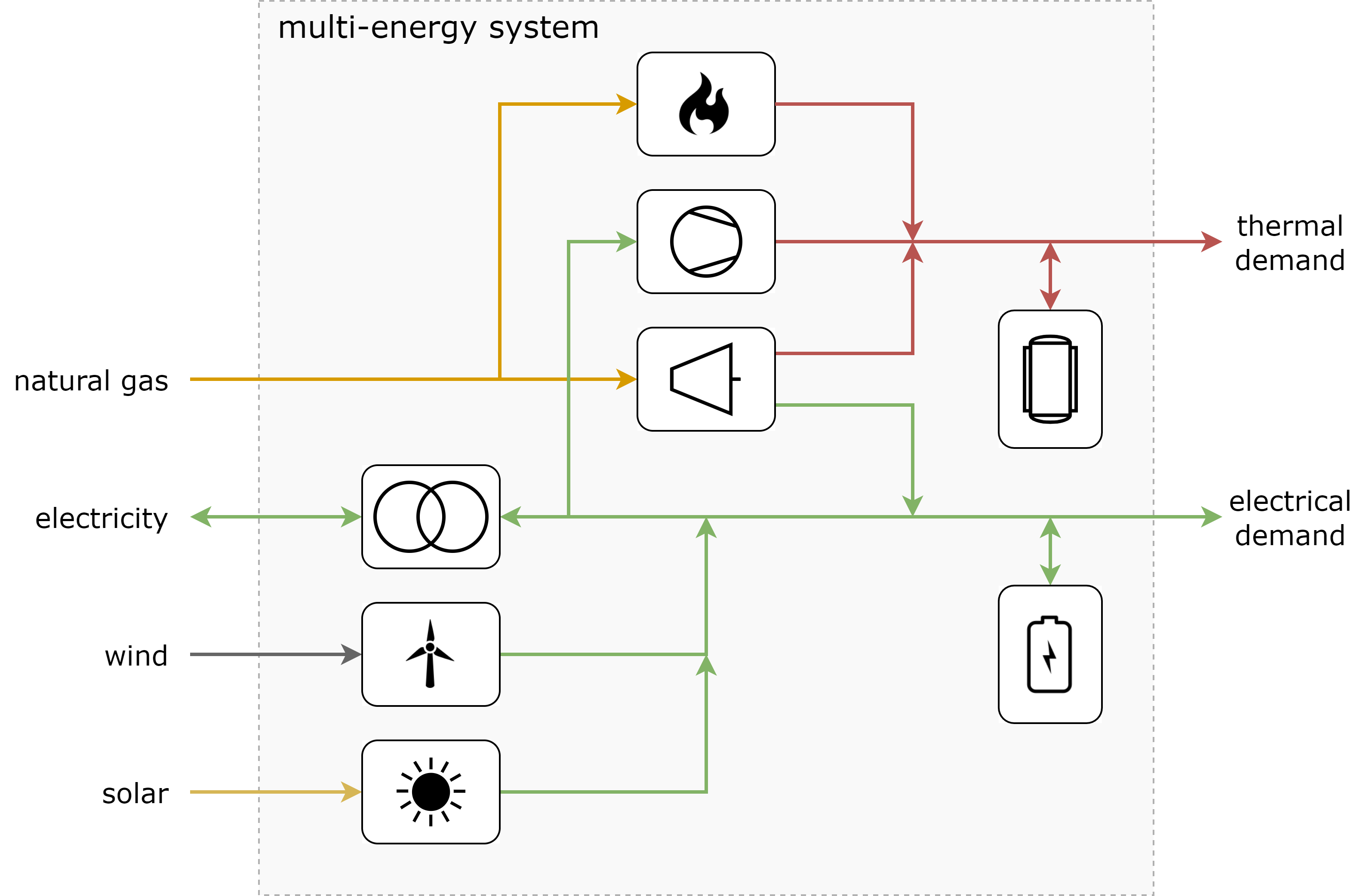}
	\caption{Topology of the multi-energy system simulation model.}
	\label{fig: case I structure}
\end{figure}

\par The energy assets, listed here in order from left to right and top to bottom, are as follows: an electric grid connection, a wind turbine, a photovoltaic (PV) installation, a natural gas boiler, a heat pump (HP), a combined heat and power (CHP) unit, a thermal energy storage system (TESS) and a battery energy storage system (BESS). The dimensions of the considered multi-energy system are summarised in \autoref{tab: mes dimensions}.

\begin{table}[!ht]
    \centering
    \begin{tabular}{c|c|c|c|c|c}
    \rowcolor[HTML]{efefef} 
    \textbf{Energy asset} & \textbf{Input} & \textbf{Output} & \textbf{P\textsubscript{nom}} & \textbf{P\textsubscript{min}} & \textbf{E\textsubscript{nom}} \\
    \hline
    grid connection & elec & elec & \(+\infty\) & \(-\infty\) & \\
    wind turbine & wind & elec & 0.8 MW\textsubscript{e} & 1.5 \% & \\
    solar PV & solar & elec & 1.0 MW\textsubscript{e} & 0 \% & \\
    boiler & CH\textsubscript{4} & heat & 2.0 MW\textsubscript{th} & 10 \% & \\
    heat pump & elec & heat & 1.0 MW\textsubscript{th} & 25 \% & \\
    CHP & CH\textsubscript{4} & heat & 1.0 MW\textsubscript{th} & 50 \% & \\
     & CH\textsubscript{4} & elec & 0.8 MW\textsubscript{e} & 50 \% & \\
    TESS & heat & heat & +0.5 MW\textsubscript{th} & -0.5 MW\textsubscript{th} & 3.5 MWh \\
    BESS & elec & elec & +0.5 MW\textsubscript{e} & -0.5 MW\textsubscript{e} & 2.0 MWh 
    \end{tabular}
    \caption{Dimensions used in the multi-energy system simulation model.}
    \label{tab: mes dimensions}
\end{table}

\par As also reported by \citeauthor{Ceusters2023SafeFunctions} \cite{Ceusters2023SafeFunctions}, the simulation model is a detailed system of differential-algebraic equations (2.548 equations with an equal number of variables). However, it does not include any simulated control system (e.g., including PID controllers). Although we acknowledge the simplification, we make an abstraction of this control layer for this case study. The resulting control error is approximately 5\%, as determined by \citeauthor{Ceusters2023SafeFunctions} \cite{Ceusters2023SafeFunctions} with separate simulations using a reduced discrete-time control horizon (5 seconds compared to 15 minutes, without any continuous \textit{error} handling). This assumption is the same for every energy management algorithm considered, so the comparison remains valid. We furthermore deliberately will not use exactly the same equations in the safety layer as in the more detailed simulation model, to better mimic a real multi-energy management case study that always will have a remaining modelling error (caused by unknown parameters, inaccurate equations, or assets not working according to specifications - according to \citeauthor{Drgona2020AllBuildings} \cite{Drgona2020AllBuildings}).

\subsection{Safety layer} \label{chapter4.3}
\par Starting from minimum and maximum electrical and thermal power output constraints:
\begin{subequations}
\begin{align}
    \label{equation8a}
    Q^{min}_{boil} \times \gamma^{t}_{boil} \leq Q^{t}_{boil} \leq Q^{max}_{boil} \times \gamma^{t}_{boil} &&\forall t&, \hspace{5pt} \gamma^{t}_{boil} \in \{0,1\} \\
    \label{equation8b}
    \begin{bmatrix}
    Q^{min}_{hp} \\
    P^{min}_{hp}
    \end{bmatrix}  \times \gamma^{t}_{hp} \leq
    \begin{bmatrix}
    Q^{t}_{hp} \\
    P^{t}_{hp}
    \end{bmatrix} \leq 
    \begin{bmatrix}
    Q^{max}_{hp} \\
    P^{max}_{hp}
    \end{bmatrix}  \times \gamma^{t}_{hp} &&\forall t&, \hspace{5pt} \gamma^{t}_{hp} \in \{0,1\} \\
    \label{equation8c}
    \begin{bmatrix}
    Q^{min}_{chp} \\
    P^{min}_{chp}
    \end{bmatrix}  \times \gamma^{t}_{chp} \leq
    \begin{bmatrix}
    Q^{t}_{chp} \\
    P^{t}_{chp}
    \end{bmatrix} \leq 
    \begin{bmatrix}
    Q^{max}_{chp} \\
    P^{max}_{chp}
    \end{bmatrix}  \times \gamma^{t}_{chp} &&\forall t&, \hspace{5pt} \gamma^{t}_{chp} \in \{0,1\} \\
    \label{equation8d}
    Q^{min}_{tess} \leq Q^{t}_{tess} \leq Q^{max}_{tess}  &&\forall t& \\
    \label{equation8e}
    P^{min}_{bess} \leq P^{t}_{bess} \leq P^{max}_{bess}  &&\forall t&
\end{align}
\end{subequations}
where \(Q^t\) and \(P^t\) are the thermal and electrical powers respectively (in accordance with \autoref{tab: mes dimensions}) and \(\gamma^{t}\) binary variables that turn on/off the given asset (i.e., as their minimal powers are not zero). Furthermore, we assume a sufficiently large grid connection so that the electrical energy balance is always fulfilled. By doing so, we focus on satisfying the thermal energy balance and no additional constraints are considered in this case study (e.g., minimal run- and downtime, ramping rates). Writing out the thermal energy balance then becomes:

\begin{subequations}
\begin{align}
    \label{equation9a}
    Q^{t}_{production} = Q^{t}_{demand} &&\forall t \\
    \label{equation9b}
    Q^{t}_{boil} + Q^{t}_{hp} + Q^{t}_{chp} + Q^{t}_{tess} = Q^{t}_{demand} && \forall t \\
    a^{t}_{boil} \cdot \eta^{t}_{boil} \cdot Q^{max}_{boil} + a^{t}_{hp} \cdot \frac{COP^{t}_{hp}}{COP^{max}_{hp}} \cdot Q^{max}_{hp} \notag\\ 
    \label{equation9c}
    + a^{t}_{chp} \cdot \eta^{t,th}_{chp} \cdot Q^{max}_{chp} + a^{t}_{tess} \cdot f(SOC^{t}_{tess}) = Q^{t}_{demand} && \forall t \\
    a^{t}_{boil} \cdot f(T^t_{boil}) \cdot Q^{max}_{boil} + a^{t}_{hp} \cdot \frac{f(T^t_{evap}, T^t_{cond})}{COP^{max}_{hp}} \cdot Q^{max}_{hp} \notag\\
    \label{equation9d}
    + a^{t}_{chp} \cdot f(P^t_{chp},Q^t_{chp},T^t_{env}) \cdot Q^{max}_{chp} 
    + a^{t}_{tess} \cdot f(\overline{T^{t}_{tess}}) = Q^{t}_{demand} && \forall t
\end{align}
\end{subequations}
where \(a\) are the (control) actions, \(\eta\) the energy efficiencies, \(COP\) the coefficient of performance, \(SOC\) the state of charge, and \(T\) various specific temperatures (i.e., \(T^t_{boil}\) the return temperature to the boiler, \(T^t_{evap}\) the evaporator temperature of the heat pump, \(T^t_{cond}\) the condenser temperature of the heat pump, \(T^t_{env}\) the environmental air temperature and \(\overline{T^{t}_{tess}}\) the average temperature in the stratified hot water storage tank). However, the different functions \(f(\cdot)\) from \autoref{equation9d} are typically not trivial to \textit{model} accurately. Our nominal models \( \Bar{(\cdot)} \) in \autoref{equation7}, which are also those used in \autoref{equation5b}, have the following metrics:

\begin{table}[H]
    \centering
    \begin{tabular}{c|c|c|c}
    \rowcolor[HTML]{efefef} 
    \textbf{Energy asset} & \textbf{R2-score} & \textbf{MAE} & \textbf{NMAE} \\
    \hline
    boiler & 99.70\% & 20.00 kW & 0.90\% \\
    heat pump & 97.05\% & 36.04 kW & 3.64\% \\
    CHP & 99.77\% & 4.20 kW & 0.35\% \\
    TESS  & 85.64\% & 39.52 kW & 4.18\% \\
    \end{tabular}
    \caption{Safety layer: nominal modelling metrics. Mean Absolute Error (MAE), Normalised Mean Absolute Error (NMAE) by range, i.e., NMAE = MAE / range(actual values)}
    \label{tab: safety layer model metrics}
\end{table}

\par In our nominal models, we have assumed a linear time-invariant relationship between the (control) action and the thermal power for the boiler and the CHP, yet a third-degree polynomial for the TESS and a second-degree polynomial for the heat pump – both also time-invariant. We note from \autoref{tab: safety layer model metrics} that the main improvement can be made in the heat pump and thermal energy storage functions, and therefore we have only included an adaptive component for these two assets:

\begin{subequations}
\begin{align}
    \label{equation10a}
    &\Hat{Q}_{hp}^t = f(a_{hp}^t, Q_{hp}^{t-1}) \\
    \label{equation10b}
    &\Hat{Q}_{tess}^t = f(a_{tess}^t, SOC_{tess}^t, Q_{tess}^{t-1}, Q_{demand}^{t-1})
\end{align}
\end{subequations}

where \(Q_{hp}^{t-1}\) is the thermal power of the heat pump, \(Q_{tess}^{t-1}\) of the thermal energy storage system, and \(Q_{demand}^{t-1}\) of the thermal demand, all one time step before (i.e., historical lag features, as is common practice due to high autocorrelation). We use Multi-layer Perceptron (MLP) regressors from \texttt{Scikit-learn} \cite{PedregosaFABIANPEDREGOSA2011Scikit-learn:Python} with ([15,10,10,10]) as the architecture of hidden layers for \autoref{equation10a} and \([25,20,20,10]\) for \autoref{equation10b} all with ReLU activation functions, using the Adam \cite{Kingma2014Adam:Optimization} solver and an adaptive learning rate. The total modelling metrics (nominal + unknown component) are given in \autoref{fig: constraint models}, using a training interval \(h_{train}\) of 1 week (672 time steps) in the first month and a monthly interval (2,688 time steps) thereafter. 

\subsection{Safe fallback policy} \label{chapter4.4}

\par The \textit{a priori} safe fallback policy \(\pi^{safe}\), can be \textit{any} (non-optimal) policy that satisfies the constraints and can typically be provided by domain experts. In our case study, this is a simple priority/cascade rule, as in \citeauthor{Ceusters2023SafeFunctions} \cite{Ceusters2023SafeFunctions} and given by \autoref{policy: safe fallback}. Note that for clarity concerns, it is written in terms of thermal power output, yet it is still converted to actions \(a^{t}_{chp}\) and \(a^{t}_{boil}\) as going from \autoref{equation9b} to \autoref{equation9c}. \vspace{5pt}

\begin{center}
\begin{minipage}{0.5\textwidth}
\begin{algorithm}[H]
\DontPrintSemicolon
\SetAlgoLined
 \eIf{\(Q^{t}_{demand} < Q^{min}_{chp}\)}{\(Q^{t}_{chp} = 0\) \\ \(Q^{t}_{boil} = Q^{t}_{demand}\)}{\eIf{\(Q^{t}_{demand} < Q^{max}_{chp}\)}{\(Q^{t}_{chp} = Q^{t}_{demand}\) \\ \(Q^{t}_{boil} = 0\)}{\(Q^{t}_{chp} = Q^{max}_{chp}\) \\ \(Q^{t}_{boil} = Q^{t}_{demand} - Q^{max}_{chp}\)}}
 \caption{safe fallback policy \cite{Ceusters2023SafeFunctions}} \label{policy: safe fallback}
\end{algorithm} \vspace{5pt}
\end{minipage}
\end{center}

\subsection{Energy managing RL agent}
\par We formulate the energy managing RL agent as a fully observable discrete-time Markov Decision Process (MDP) with the tuple \( \langle S,A,P_a,R_a \rangle \) so that:

\begin{subequations}
\begin{align}
    \label{equation11a}
    S^t = ( E_{th}^t,\ E_{el}^t,\ P_{wind}^t,\ P_{solar}^t,\ X_{el}^t, \ SOC_{tess}^t, \ SOC_{bess}^t, \ H^t, \ D^t  )& & S^t \in S\\[1em]
    \label{equation11b}
    a_{boil}^t = (0,\ a_{boil}^{min} \xrightarrow{} a_{boil}^{max})& & a_{boil}^t \in A \\
    \label{equation11c}
    a_{hp}^t = (0,\ a_{hp}^{min} \xrightarrow{} a_{hp}^{max})& & a_{hp}^t \in A \\
    \label{equation11d}
    a_{chp}^t = (0,\ a_{chp}^{min} \xrightarrow{} a_{chp}^{max})& & a_{chp}^t \in A \\
    \label{equation11e}
    a_{tess}^t = (a_{tess}^{min} \xrightarrow{} a_{tess}^{max})& & a_{tess}^t \in A \\
    \label{equation11f}
    a_{bess}^t = (a_{bess}^{min} \xrightarrow{} a_{bess}^{max})& & a_{bess}^t \in A \\[1em]
    \label{equation11g}
    P_a(s, s') = Pr(s_{t+1} = s' \ | \ s_t = s, a_t = a)& \\
    \label{equation11h}
    R_a(s, s') = - (x \times L_{cost}^t + y \times L_{comfort}^t) - z&
\end{align}
\end{subequations}
where \( E_{th}^t \) is the thermal demand, \( E_{el}^t \) the electrical demand, \( P_{wind}^t \) the electrical wind in-feed, \( P_{solar}^t \) the electrical solar in-feed, \(X_{el}^t\) the electrical price signal, \(SOC_{tess}^t\) the state-of-charge (SOC) of the TESS, \(SOC_{bess}^t\) the SOC of the BESS, \(H^t\) the hour of the day and \(D^t\) the day of the week all at the \textit{t}-th step, which constitute the state-space \( S \). The action space \( A \) includes the control set-points from, \( a_{boil}^t \) the natural gas boiler, \( a_{hp}^t \) the heat pump, \( a_{chp}^t \) the CHP unit,  \( a_{tess}^t \) the TESS, and \( a_{bess}^t \) the BESS all between the minimum and maximum power rates in accordance with \autoref{tab: mes dimensions}. Moreover, \( P_a \) signifies a transition probability, that exclusively relies on the current state and is unaffected by prior states (in other words, adhering to the Markov Property), for when the system is in a specific state \( s \in S \) at time step \( t \) and takes action \( a \in A \), which would result in the system being in state \( s' \in S \) at the subsequent time step \( t+1 \).

\par We formulate the objective, that is, the reward function, so that when maximising this function (via \autoref{equation4a}) we minimise the positive version of that function. Hence, we minimise the energy costs \( L_{cost}^t \) in EUR and the loss in (thermal) comfort \( L_{comfort}^t = | Q^{t}_{demand} - Q^{t}_{production} | \) in Watt with scalarisation weights \( x = 1/10 \) and \( y = 1/5e5\) and with an additional cost \( z = 1 \) to further shape the reward when the original, uncorrected, predicted action \(\tilde a\) was \textit{expected} to violate constraints. The loss in (thermal) comfort, \( L_{comfort}^t \), serves as an additional fine-tuning mechanism to further mitigate the modelling error of the constraints itself (see \autoref{tab: safety layer model metrics}), i.e., to further shape the reward and thus guide the agent towards safer actions. The discrete-time control horizon is 15 minutes. The state-space \( S \) is normalised and all actions in the action-space \( A \) are scaled between \( [+1, -1] \).

\par As the specific RL algorithm, we use a twin delayed deep deterministic policy gradient agent (TD3, that is, twin delayed DDPG) from the \texttt{ stable baseline} \cite{stable-baselines3} implementations. The pseudocode and the used hyperparameters of the TD3 algorithm are given in \hyperref[Appendix C]{Appendix~\ref*{Appendix C}}.

\subsection{Evaluation}
\par We evaluate our methods, \texttt{OptLayerPolicy} and \texttt{GreyOptLayerPolicy}, against a \textit{vanilla} RL agent (i.e., without a safety layer and therefore being unsafe), the original \texttt{OptLayer} from \citeauthor{Pham2018OptLayerWorld} \cite{Pham2018OptLayerWorld} and the original \texttt{SafeFallback} from \citeauthor{Ceusters2023SafeFunctions} \cite{Ceusters2023SafeFunctions} (both as state-of-the-art benchmarks), in a week-long evaluation environment while having a separate year-long training environment using the simulation model as discussed in \autoref{chapter4.2}. This is in terms of reducing energy costs subject to compliance with constraints (i.e., all RL agents have the same \autoref{equation11h}). We also include random agents to further study the effectiveness of the safety layers and to serve as a minimal learning benchmark. The linear MPC from \citeauthor{Ceusters2021Model-predictiveStudies} \cite{Ceusters2021Model-predictiveStudies} is not included here, as the constraints can be formulated directly in the method. Note that any uncertainty (e.g., from demands, prices, or renewables) is inherently handled by the RL agents in accordance with the \textit{expectation} \( E\{ \cdot \} \) in \autoref{equation4a}. The experiments are carried out on a local machine that has an Intel® Core™ i5-8365U CPU  @1.6GHz, 16 GB of RAM, and an SSD.

\section{Results and discussion} \label{chapter5}
\par
\begin{figure}[H]
    \centering
    \includegraphics[width=0.90\textwidth, height=0.4\textheight]{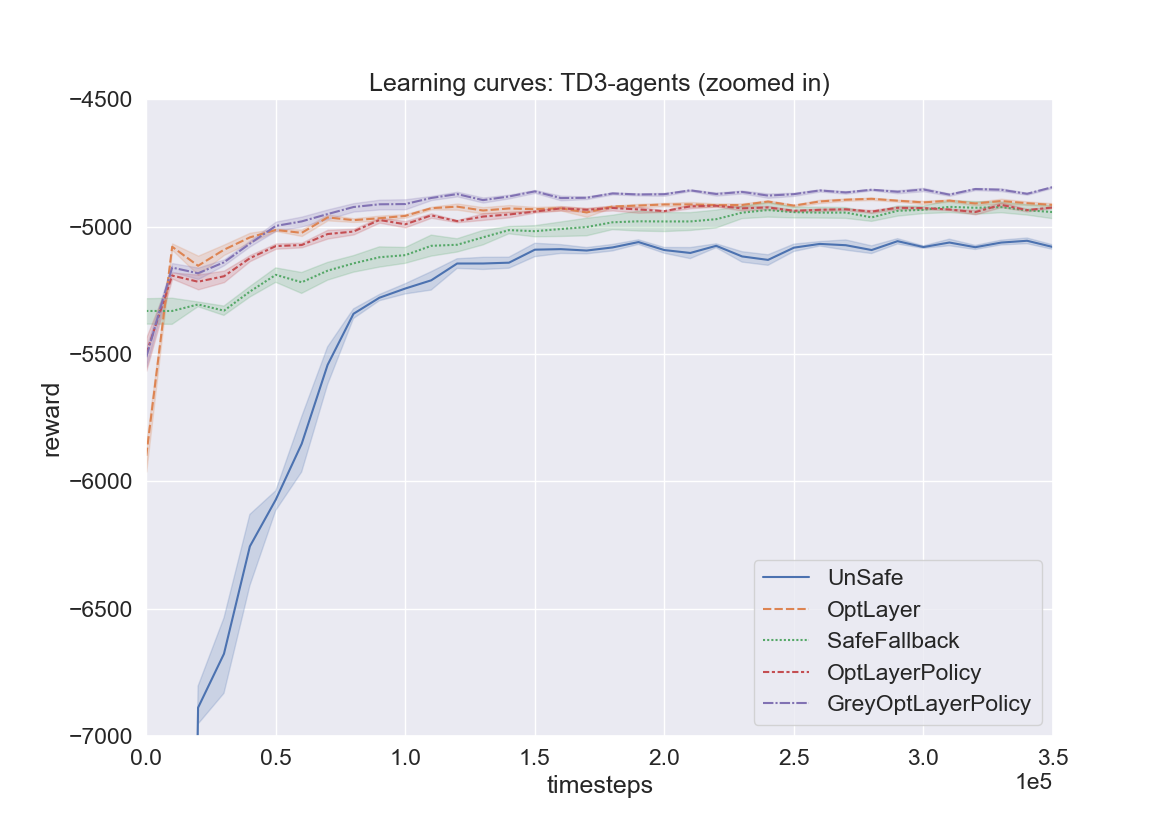}
    \caption{5-run average learning curves with a training budget of 10 years worth of time steps per run (i.e., 350,400 time steps per run). Note that the y-axis is zoomed in (see \autoref{fig: learning curve zoomed out} in \hyperref[Appendix B]{Appendix~\ref*{Appendix B}} for the zoomed-out version)}
    \label{fig: learning curve}
\end{figure}

\begin{figure}[H]
    \centering
    \includegraphics[width=0.90\textwidth, height=0.4\textheight]{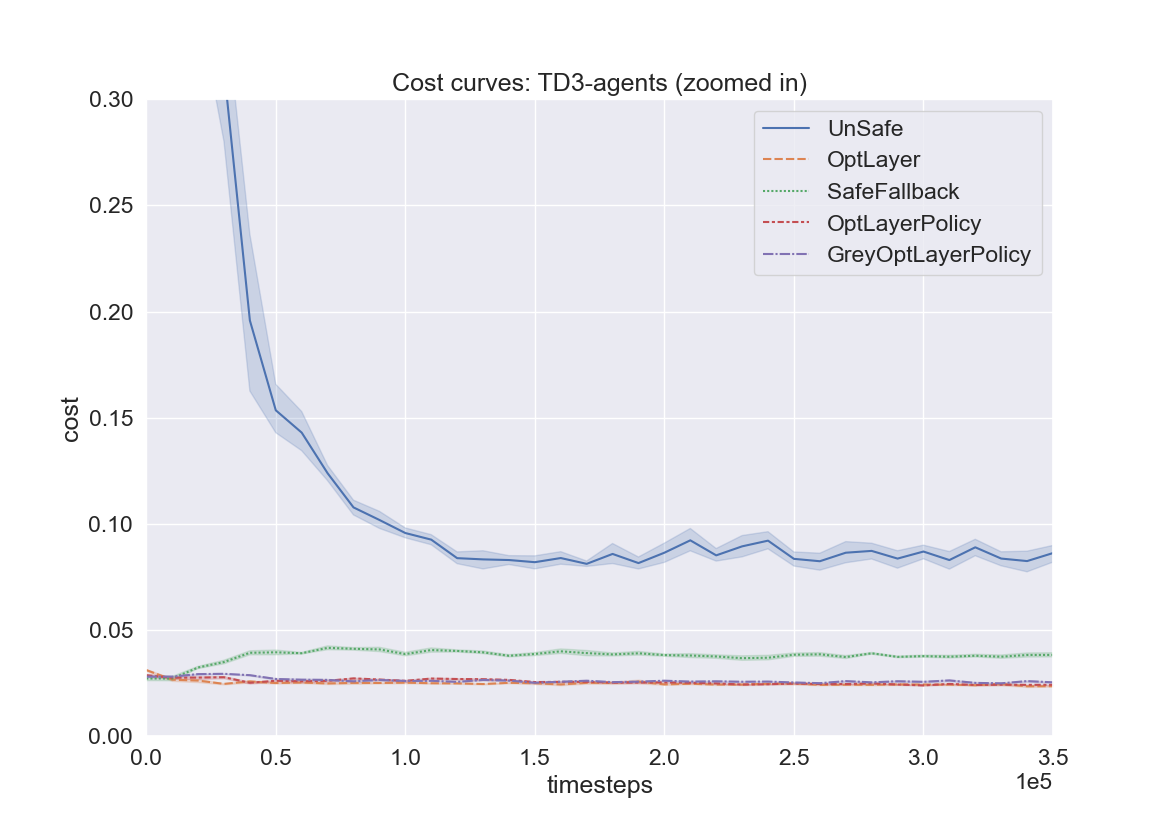}
    \caption{5-run average cost curve (i.e., normalised mean absolute error) with a training budget of 10 years worth of time steps per run (i.e., 350,400 time steps per run). Note that the y-axis is zoomed in (see \autoref{fig: cost curve zoomed out} in \hyperref[Appendix B]{Appendix~\ref*{Appendix B}} for the zoomed-out version).}
    \label{fig: cost curve}
\end{figure}

\begin{figure}[H]
    \centering
    \begin{subfigure}{0.5\textwidth}
        \begin{minipage}{0.9\textwidth}
            \includegraphics[width=\textwidth]{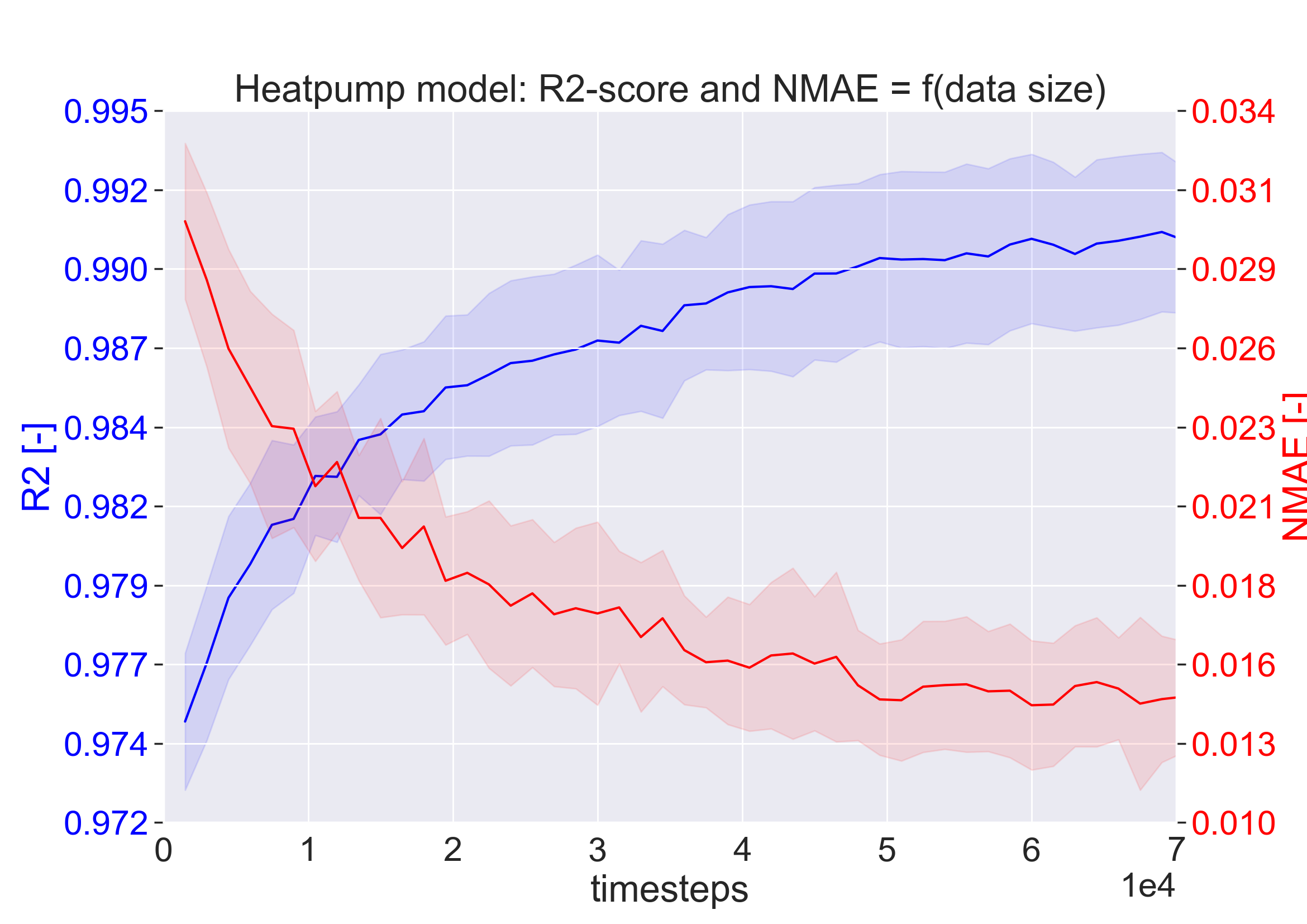}
            \caption{Heat pump: \( \Hat{Q}_{hp}^t = f(a_{hp}^t, Q_{hp}^{t-1}) \) with [25,20,20,10] hidden neurons using ReLU activation functions}
        \end{minipage}
        \label{fig: constraint model hp}
    \end{subfigure}%
    \begin{subfigure}{0.5\textwidth}
        \begin{minipage}{0.9\textwidth}
            \includegraphics[width=\textwidth]{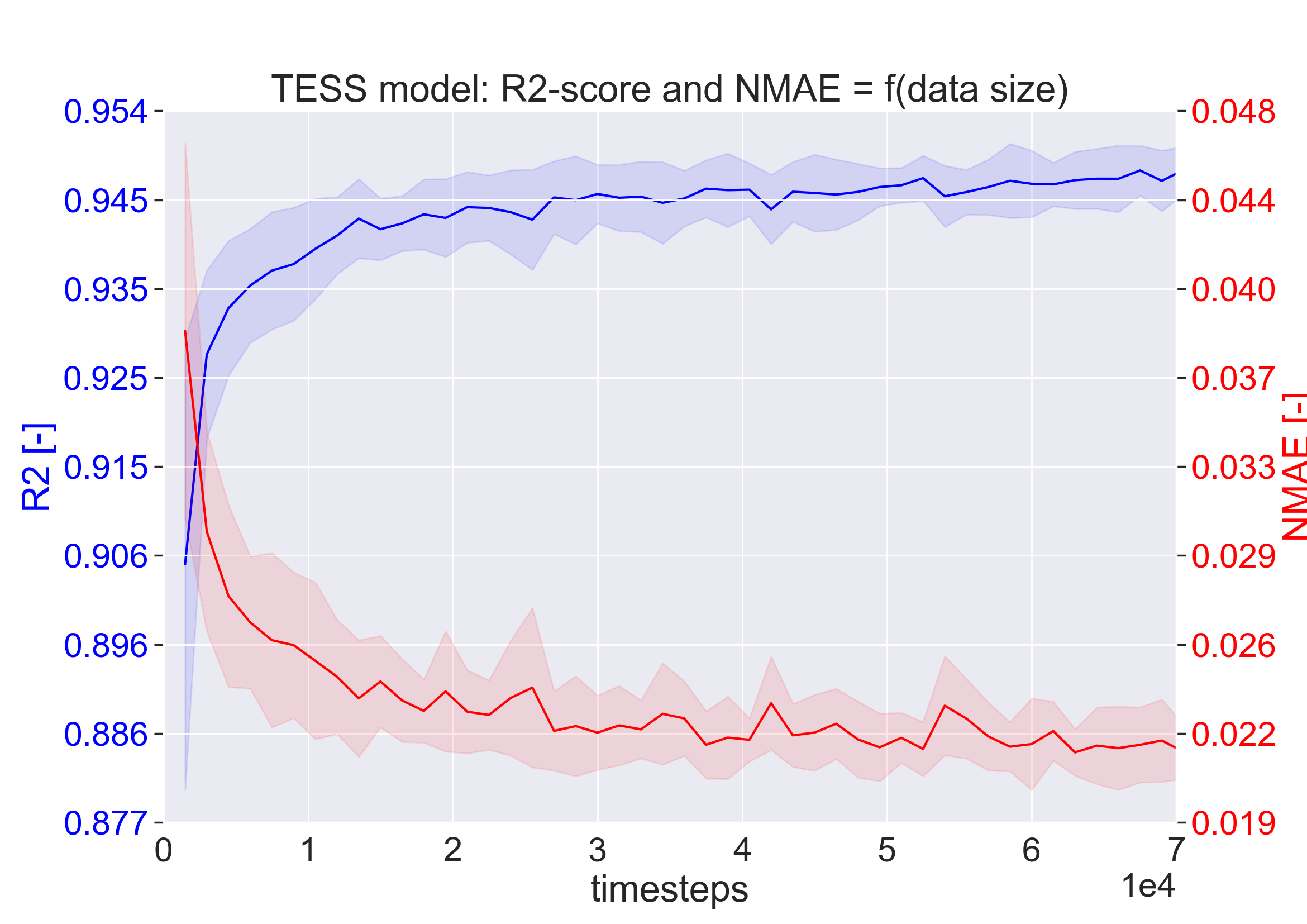}
            \caption{Thermal energy storage system: \( \Hat{Q}_{tess}^t = f(a_{tess}^t, SOC_{tess}^t, Q_{tess}^{t-1}, Q_{demand}^{t-1}) \) with [15,10,10,10] hidden neurons using ReLU activation functions}
        \end{minipage}
        \label{fig: constraint model tess}
    \end{subfigure}
    \vspace*{10pt}
    \caption{Safety layer: total (nominal and unknown component, i.e. \autoref{equation7}) modelling metrics for the first 2 years' worth of data, while interacting with the environment. The Mean Absolute Error is normalised by range (right-hand y-axis).}
    \label{fig: constraint models}
\end{figure}

 The simulated performance, in terms of minimisation of energy costs subject to the fulfilment of the constraints, is shown in \autoref{fig: learning curve} and \autoref{fig: cost curve}, with their numerical values after training (at time step 350,400) in \autoref{tab: training results} and before training (at time step 0) in \autoref{tab: random results}. The objective values (i.e., the rewards using \autoref{equation11h}, which has both \( L_{cost}^t \) and \( L_{comfort}^t \) ) are shown both in absolute values and relative to the \textit{vanilla} RL benchmark (i.e., without a safety layer). The constraint tolerance we define here as the difference between the \textit{true} constraints (i.e., observed after executing the control actions in the more detailed simulation environment) and the \textit{modelled} constraints in the safety layer (i.e., computed in the safety layer itself, which is only \textit{believed} to be true – see \autoref{tab: safety layer model metrics} and \autoref{fig: constraint models}). While, in principle, this is a calculation for the complete constraint set, we focus here on the most limiting constraint being the thermal energy balance of \autoref{equation9d}. Note that, this is only for the constraint tolerance metric and that \autoref{equation8a} till \autoref{equation8e} are still part of the mixed-integer quadratic program in the safety layer (this as their influence on the metric over the complete constraint set is marginal and can therefore be neglected). The constraint tolerance metric is then shown as the mean average error normalised to the thermal demand range (NMAE) and shown as the normalised sum (NSUM) over all evaluation time steps (i.e., 0\% would then mean perfect thermal energy balance and thus constraint fulfilment, over all evaluation time steps). We again want to emphasise the lack of a simulated control system (as discussed in \autoref{chapter4.2}), resulting in an NSUM of approximately 5\%, as determined by \citeauthor{Ceusters2023SafeFunctions} \cite{Ceusters2023SafeFunctions} in the same simulated case study, and the deliberately more detailed simulation model itself (to better mimic a real case study).

\begin{table}[H]
    \centering
    \begin{tabular}{l|c|c|c|c}
    \rowcolor[HTML]{EFEFEF} 
    \multicolumn{1}{l|}{\cellcolor[HTML]{EFEFEF}}                                              & \multicolumn{2}{c|}{\cellcolor[HTML]{EFEFEF}\textbf{Objective value}}                                         & \multicolumn{2}{c}{\cellcolor[HTML]{EFEFEF}\textbf{Constraint tolerance}} \\
    \rowcolor[HTML]{EFEFEF} 
    \multicolumn{1}{l|}{\multirow{-2}{*}{\cellcolor[HTML]{EFEFEF}\textbf{EMS algorithm}}} & \multicolumn{1}{c|}{\cellcolor[HTML]{EFEFEF}absolute} & \multicolumn{1}{c|}{\cellcolor[HTML]{EFEFEF}relative} & \multicolumn{1}{c|}{\cellcolor[HTML]{EFEFEF}NMAE} & \multicolumn{1}{c}{\cellcolor[HTML]{EFEFEF}NSUM}   \\
    \hline
    Unsafe TD3 & -5,080 & 100.0\% & 8.1\% & 22.3\% \\
    OptLayer \cite{Pham2018OptLayerWorld} TD3 & -4,915 & 103.4\% & \textbf{2.3\%} & \textbf{6.1\%} \\
    SafeFallback \cite{Ceusters2023SafeFunctions} TD3 & -4,943 & 102.8\% & 3.9\% & 9.9\% \\ 
    OptLayerPolicy TD3 & -4,924 & 103.2\% & 2.4\% & 6.2\% \\
    GreyOptLayerPolicy TD3 & \textbf{-4,844} & \textbf{104.9\%} & 2.5\% & 6.5\% \\
    \end{tabular}
    \caption{5-run average policy performance with a training budget of 10 years worth of time steps per run (i.e., 350,400 time steps per run)}
    \label{tab: training results}
\end{table}

\par These results show that our \texttt{GreyOptLayerPolicy} method outperforms all other benchmarks after training (104.9\%), yet with a slightly worse constraint tolerance (2.5\%) than the original \texttt{OptLayer} method (2.3\%). Also, the initial utility of both of our proposed methods is significantly higher (92.4\% and 92.2\%) compared to \texttt{OptLayer} (86.1\%), as initially, they use the \texttt{SafeFallback} policy \( \pi^{safe} \) itself (97.2\%) – due to exceeding the set threshold \( h_{safe} \) in \autoref{algo: optlayerpolicy}. All safety layer methods are, as intended, significantly safer than the unconstrained \textit{vanilla} TD3 benchmark, which has an initial NMAE of 56.5\% and reaches 8.1\% under the consequence of the \( L_{comfort}^t \) term in \autoref{equation11h}. The higher constraint tolerance of the original \texttt{SafeFallback} method itself can be explained by the fact that it is not capable of handling equality constraints \cite{Ceusters2023SafeFunctions} (as otherwise the constraint \textit{check} would seldom be passed). The constraint tolerances of the other safety layer methods are of the same order of magnitude, both before and after training – all in line with the NMAE of the constraint functions themselves (\autoref{tab: safety layer model metrics} and \autoref{fig: constraint models}). The small difference before training between \texttt{OptLayerPolicy} and \texttt{GreyOptLayerPolicy} can be explained by the remaining variance in the results, as at this stage they both use the (same) nominal constraints.

\begin{table}[H]
    \centering
    \begin{tabular}{l|c|c|c|c}
    \rowcolor[HTML]{EFEFEF} 
    \multicolumn{1}{l|}{\cellcolor[HTML]{EFEFEF}}                                              & \multicolumn{2}{c|}{\cellcolor[HTML]{EFEFEF}\textbf{Objective value}}                                         & \multicolumn{2}{c}{\cellcolor[HTML]{EFEFEF}\textbf{Constraint tolerance}} \\
    \rowcolor[HTML]{EFEFEF} 
    \multicolumn{1}{l|}{\multirow{-2}{*}{\cellcolor[HTML]{EFEFEF}\textbf{EMS algorithm}}} & \multicolumn{1}{c|}{\cellcolor[HTML]{EFEFEF}absolute} & \multicolumn{1}{c|}{\cellcolor[HTML]{EFEFEF}relative} & \multicolumn{1}{c|}{\cellcolor[HTML]{EFEFEF}NMAE} & \multicolumn{1}{c}{\cellcolor[HTML]{EFEFEF}NSUM} \\
    \hline
    Unsafe Random & -14,223 & 35.7\% & 56.5\% & 146.0\% \\
    OptLayer \cite{Pham2018OptLayerWorld} Random & -5,900 & 86.1\% & 3.1\% & 8.0\% \\
    SafeFallback \cite{Ceusters2023SafeFunctions} Random & -5,331 & 95.3\% & 2.7\% & 7.0\% \\
    SafeFallback \cite{Ceusters2023SafeFunctions} (\(\pi^{safe}\)) & \textbf{-5,228} & \textbf{97.2\%} & \textbf{2.4\%} & \textbf{6.3\%} \\
    OptLayerPolicy Random & -5,499 & 92.4\% & 2.9\% & 7.4\% \\
    GreyOptLayerPolicy Random & -5,512 & 92.2\% & 2.8\% & 7.2\% \\
    \end{tabular}
    \caption{5-run average policy performance before any training (i.e., the initial utility)}
    \label{tab: random results}
\end{table}

\par In the learning curves of the TD3 agents (\autoref{fig: learning curve}) we observe a steep initial learning rate (except for the original \texttt{SafeFallback} method), low variance and stable performance with an increasing amount of interactions with its environment. As already reported here, we observe that our \texttt{OptLayerPolicy} and \texttt{GreyOptLayerPolicy} have a significantly higher initial utility compared to the original \texttt{OptLayer} and \textit{vanilla} Unsafe approach. This, again, is due to the additional \textit{a priori} expert knowledge in the form of the safe fallback policy \( \pi^{safe} \) and in the form of the constraint functions themselves (which is also true for the \texttt{OptLayer} approach compared to the unsafe \textit{vanilla} RL agent). The \texttt{GreyOptLayerPolicy} algorithm surpasses \texttt{OptLayerPolicy} just after \( \approx \) 20,000 time steps (7 months) and \texttt{OptLayer} after \( \approx \) 50,000 time steps (17 months). The original \texttt{SafeFallback} method gets quickly surpassed by all other safety layer methods (all around 5,000 time steps, e.g. after 5,155 time steps for the \texttt{GreyOptLayerPolicy} method).

\par In the cost curves of the TD3 agents (\autoref{fig: cost curve}) we observe a relatively constant constraint tolerance when using a safety layer (although, in the \texttt{GreyOptLayerPolicy} method, the constraint functions themselves also improve – as can be seen in \autoref{fig: constraint models} and compared to \autoref{tab: safety layer model metrics}, i.e. starting from 3.64\% NMAE for the heat pump to 1.14\% and from 4.18\% for the thermal energy storage system to 2.21\%), which can be explained by the fact that energy fulfilment is conflicting with the energy cost minimisation objective (i.e., without a thermal demand fulfilment constraint, all thermal production would be turned off as these technical units consume and thus cost energy). The energy-minimising RL agents, therefore, push the multi-energy system to the respective limit of their energy balancing constraints (and thus exploit the nominal and remaining modelling error of the constraint functions themselves). The unconstrained \textit{vanilla} TD3 benchmark does have a significant drop in its constraint tolerance under the consequence of the \( L_{comfort}^t \) term in \autoref{equation11h} (which has far less impact when using a safety layer, yet still results in a slightly better constraint tolerance for \texttt{OptLayer} and \texttt{OptLayerPolicy} - as its predicted actions \( \tilde a \) are more "falsely considered" to be unsafe giving the additional cost \( z \) and a higher \( L_{comfort}^t \), which essentially is an overcompensation towards safer actions). As mentioned before, the higher constraint tolerance of the original \texttt{SafeFallback} method itself can be explained by the fact that it is not capable of handling equality constraints (since otherwise the constraint \textit{check} would seldom be passed), forcing a relaxation of the thermal energy balance constraint.

\section{Conclusion} \label{chapter6}
\par This paper presented a novel model-free safe RL approach of the following \textit{combined} characteristics: (i) providing hard-constraints satisfaction guarantees (ii) while decoupled from the RL (as a Markov Decision Process) formulation, (iii) both during training a (near) optimal policy (involving exploratory and exploitative steps) as well as during the deployment of any policy (e.g., offline pre-trained RL agents) and (iv) this while learning uncertain constraint components and safely improving the policy with high sample efficiency, and (v) starting from an increased initial utility, to (vi) demonstrate for the energy management of multi-energy systems.
\par We conclude that, while special attention is required to introduce surrogate functions into the optimisation problem, the \texttt{GreyOptLayerPolicy} method is the most advantageous due to both the increase of its initial utility (92.2\% compared to 86.1\% for \texttt{OptLayer}) \textit{and} the ability to self-improve its constraints (1.14\% NMAE starting from 3.64\% for the heat pump and 2.21\% NMAE starting from 4.18\% for the thermal energy storage system), leading to better policies (104.9\% compared to 103.4\% for \texttt{OptLayer}). This is despite the initial utility for the \texttt{GreyOptLayerPolicy} method being lower than the \texttt{SafeFallback} method, as this gets quickly surpassed (after 5,155 time steps) due to the high sample efficiency of \texttt{GreyOptLayerPolicy} and also despite the general reliance on the availability of nominal constraints and a safe fallback policy. Finally, we propose the following directions for future work:
\begin{itemize}
    \item Ensuring the robustness of RL-based energy management systems against faulty and noisy measurements or observations while utilising the capabilities of online hyperparameter optimisation methods - wherein the hyperparameters of the RL agent are dynamically tuned during training.
    \item Reduce computational complexity of the safety layer when using surrogate constraint functions or parameters so that more detailed function approximation architectures can be used (e.g., more neurons and layers in an ANN).
    \item Verification of the proposed safety layers in a controlled laboratory environment, using inherently unsafe control methods (e.g., random and RL agents, as in this work).
\end{itemize} 

\section{Acknowledgement}
This research has received equal support from ABB n.v. and the Flemish Agency for Innovation and Entrepreneurship (VLAIO) under grant HBC.2019.2613.

\section*{CRediT authorship contribution statement}
\par \textbf{Glenn Ceusters}: Conceptualisation, Methodology, Software, Validation, Formal analysis, Resources, Data curation, Writing - original draft, Visualisation, Funding acquisition; \textbf{Muhammad Andy Putratama}: Conceptualisation, Writing - review and editing, Supervision; \textbf{Rüdiger Franke}: Supervision;  \textbf{Ann Nowé}: Writing - review and editing, Supervision; \textbf{Maarten Messagie}: Supervision. 

%\newpage
\appendix
\renewcommand\thefigure{\thesection.\arabic{figure}}
\renewcommand\thetable{\thesection.\arabic{table}}
\section{Time series visualisations}
\label{Appendix A}

\par
\begin{figure}[H]
    \centering
    \includegraphics[width=\textwidth,height=\textheight,keepaspectratio]{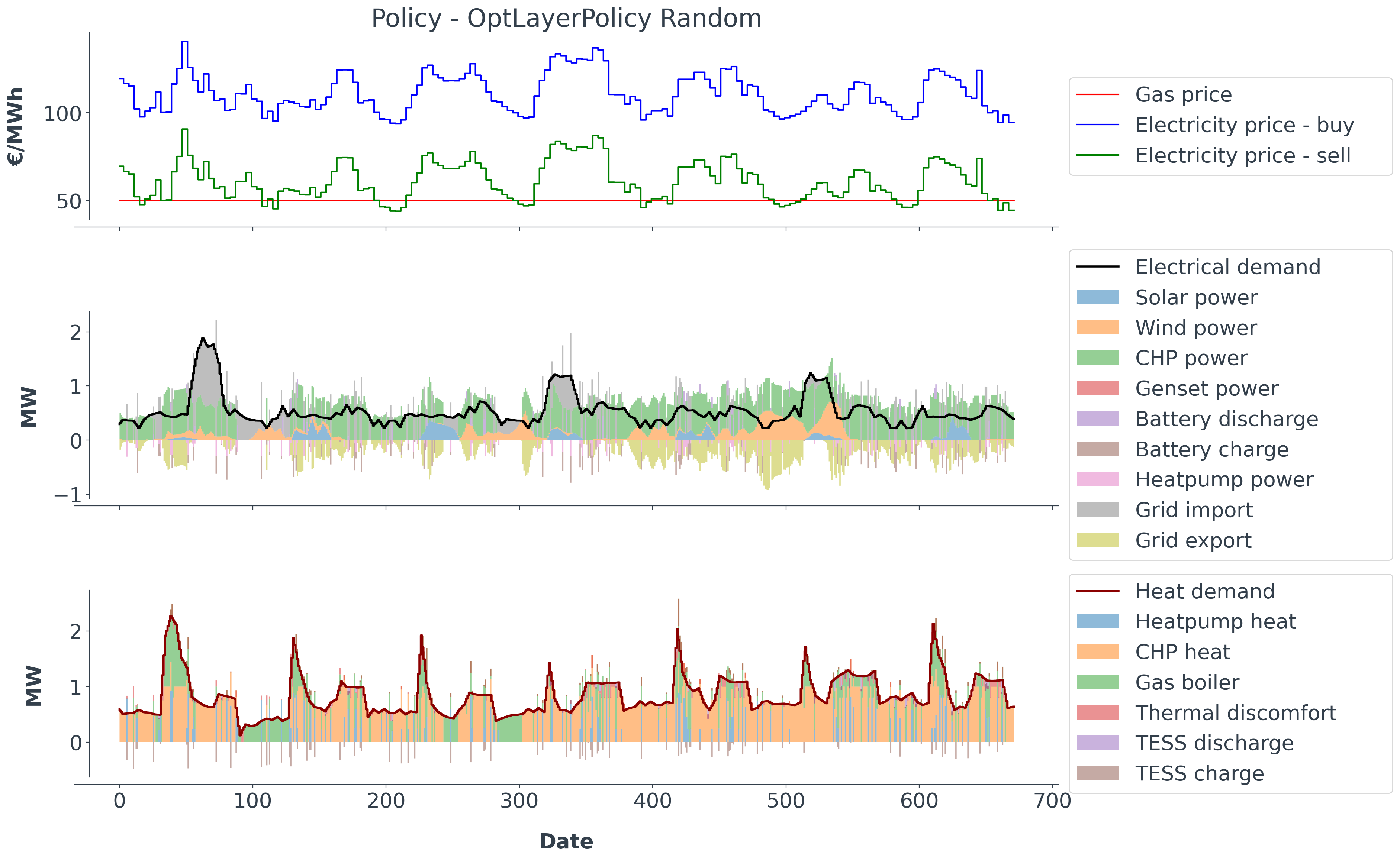}
    \caption{Policy visualisation: OptLayerPolicy random (or TD3 before training)}
    \label{fig: policy optlayerpolicy random}
\end{figure}

\begin{figure}[H]
    \centering
    \includegraphics[width=\textwidth,height=\textheight,keepaspectratio]{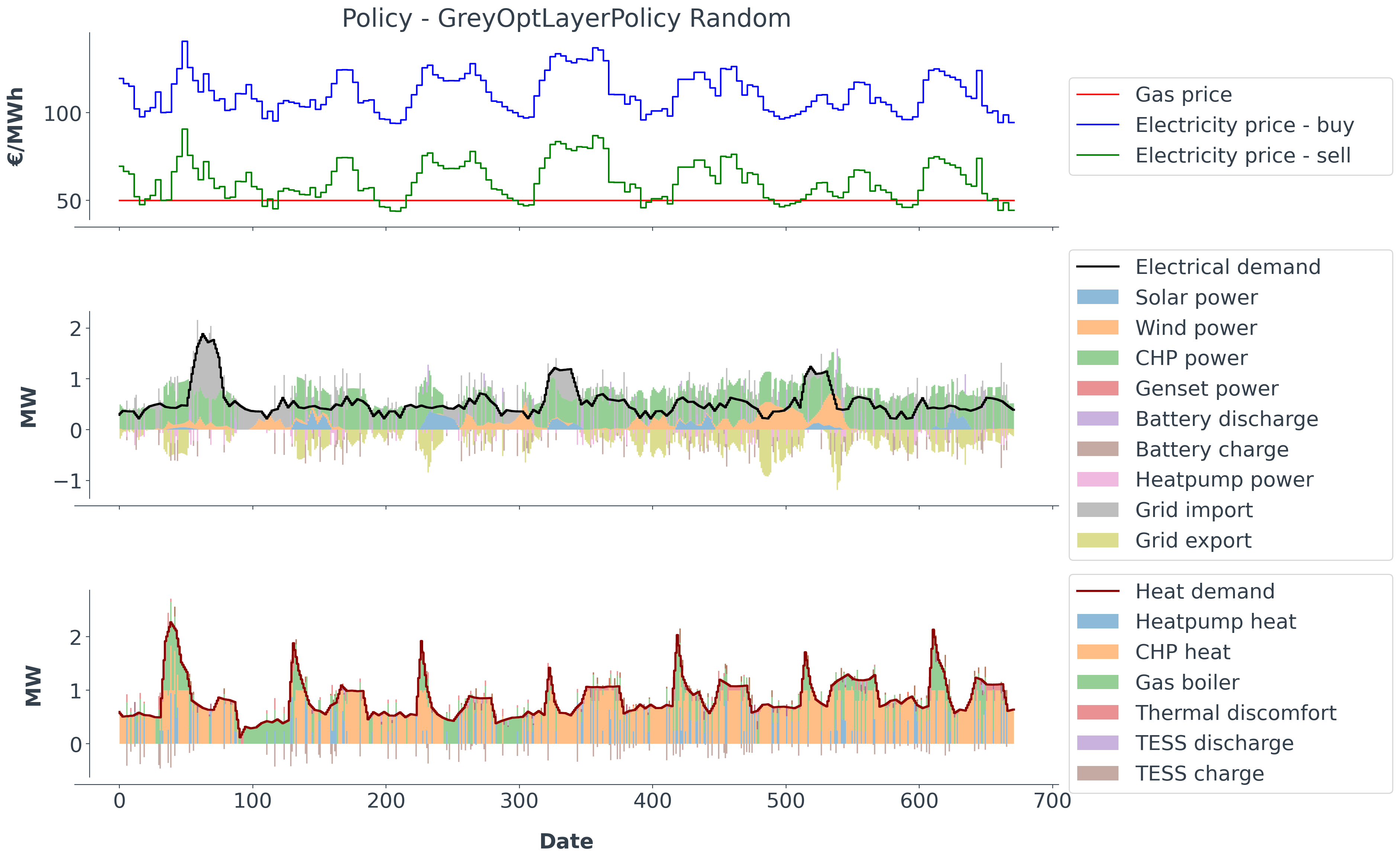}
    \caption{Policy visualisation: GreyOptLayerPolicy random (or TD3 before training)}
    \label{fig: policy greyoptlayerpolicy random}
\end{figure}

 In this first appendix, we show time series visualisation samples of 1 week using the self-discovered control policies. The plots and their description of the benchmarks (i.e., \texttt{UnSafe}, \texttt{OptLayer} and \texttt{SafeFallback}) are not repeated here, as they can be found in \citeauthor{Ceusters2023SafeFunctions} \cite{Ceusters2023SafeFunctions}. Our first observation is that the initial policy of both \texttt{OptLayerPolicy} (\autoref{fig: policy optlayerpolicy random}) and \texttt{GreyOptLayerPolicy} (\autoref{fig: policy greyoptlayerpolicy random}) are very similar. This is, at this stage, they have the same (nominal) constraints, and occasionally the predicted actions, \(\tilde a\), are close enough to the feasible solution space (calculated using \autoref{equation6}) so that they are minimally corrected and thus do not exceed the given threshold \( h_{safe} \). As intended, most of the time this threshold is exceeded, and thus the safe fallback policy \( \pi_{safe} \) is used – in accordance with \autoref{algo: optlayerpolicy}.

\par
\begin{figure}[H]
    \centering
    \includegraphics[width=\textwidth,height=\textheight,keepaspectratio]{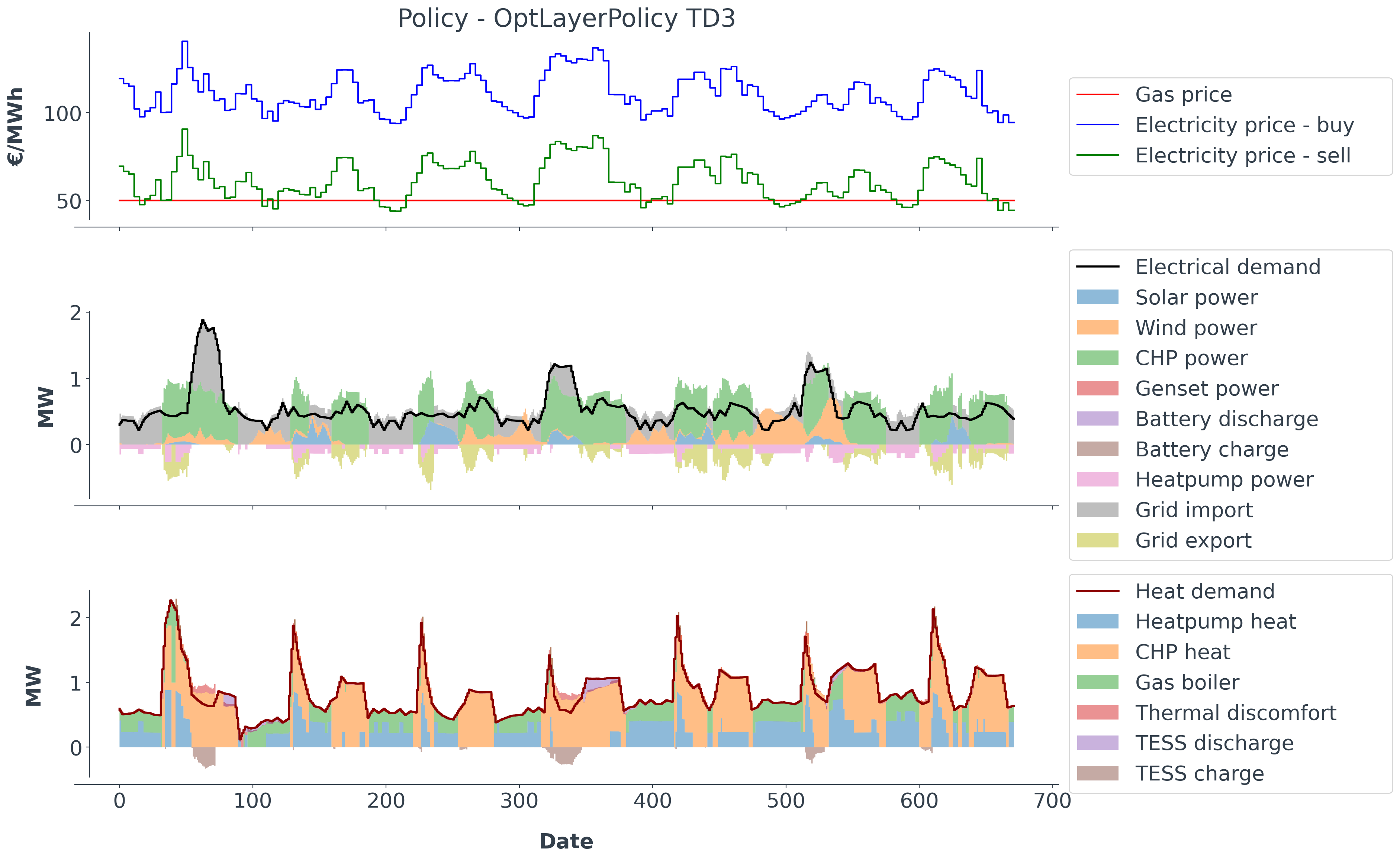}
    \caption{Policy visualisation: OptLayerPolicy TD3 (after safe training)}
    \label{fig: policy optlayerpolicy TD3}
\end{figure}

\begin{figure}[H]
    \centering
    \includegraphics[width=\textwidth,height=\textheight,keepaspectratio]{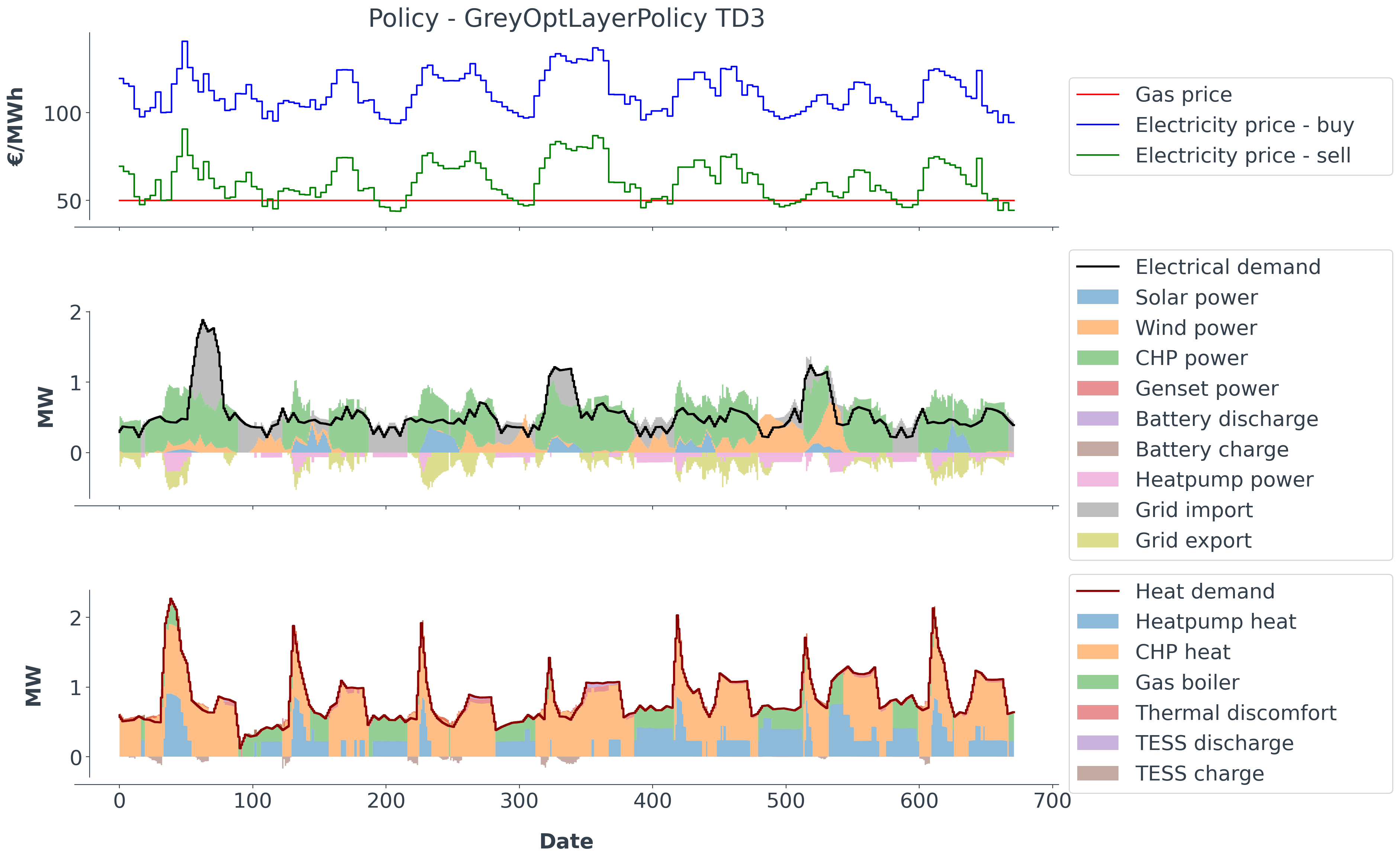}
    \caption{Policy visualisation: GreyOptLayerPolicy TD3 (after safe training)}
    \label{fig: policy greyoptlayerpolicy TD3}
\end{figure}

 When analysing the policies of the TD3 agents, after safely training them (i.e., \autoref{fig: policy optlayerpolicy TD3} and \autoref{fig: policy greyoptlayerpolicy TD3}), we observe that both continue to have a low constraint tolerance, as expected. However, the \texttt{GreyOptLayerPolicy} policy is slightly better at using the heat pump when electricity prices are low, producing electricity with the CHP and charging the TESS simultaneously when electricity prices are high, and vice versa. Hence, it is better at minimising energy costs. In both cases, however, the BESS is underutilised, i.e., it fails to learn to charge the BESS when electricity prices are low and discharging the BESS when prices are high. This underuse of the BESS was also reported in previous work (e.g., \citeauthor{Ceusters2021Model-predictiveStudies} \cite{Ceusters2021Model-predictiveStudies} and \citeauthor{Ceusters2023SafeFunctions}\cite{Ceusters2023SafeFunctions}), which will require revisiting the state-space formulation (\autoref{equation11a}) and the consideration of specific reward shaping in \autoref{equation11h}.

\section{Learning and cost curves: zoomed out}
\label{Appendix B}
\setcounter{figure}{0}   

\par This appendix shows the zoomed-out learning and cost curves of the TD3 agents so that all curves are fully visible, i.e., so that the UnSafe curves are visible for all time steps.

\begin{figure}[H]
    \centering
    \includegraphics[width=0.90\textwidth, height=0.4\textheight]{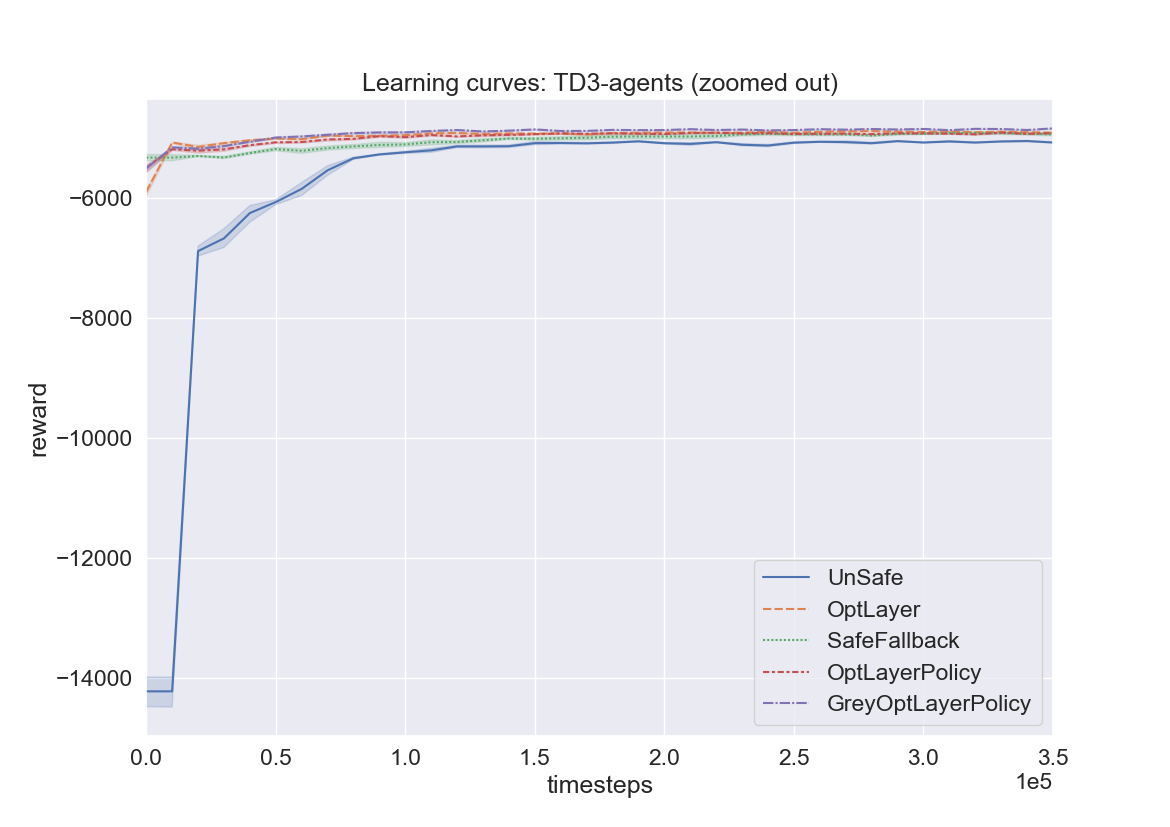}
    \caption{5-run average learning curves with a training budget of 10 years worth of time steps per run (i.e., 350,400 time steps per run). Note that the zoomed-in version is \autoref{fig: learning curve}.}
    \label{fig: learning curve zoomed out}
\end{figure}

\begin{figure}[H]
    \centering
    \includegraphics[width=0.90\textwidth, height=0.4\textheight]{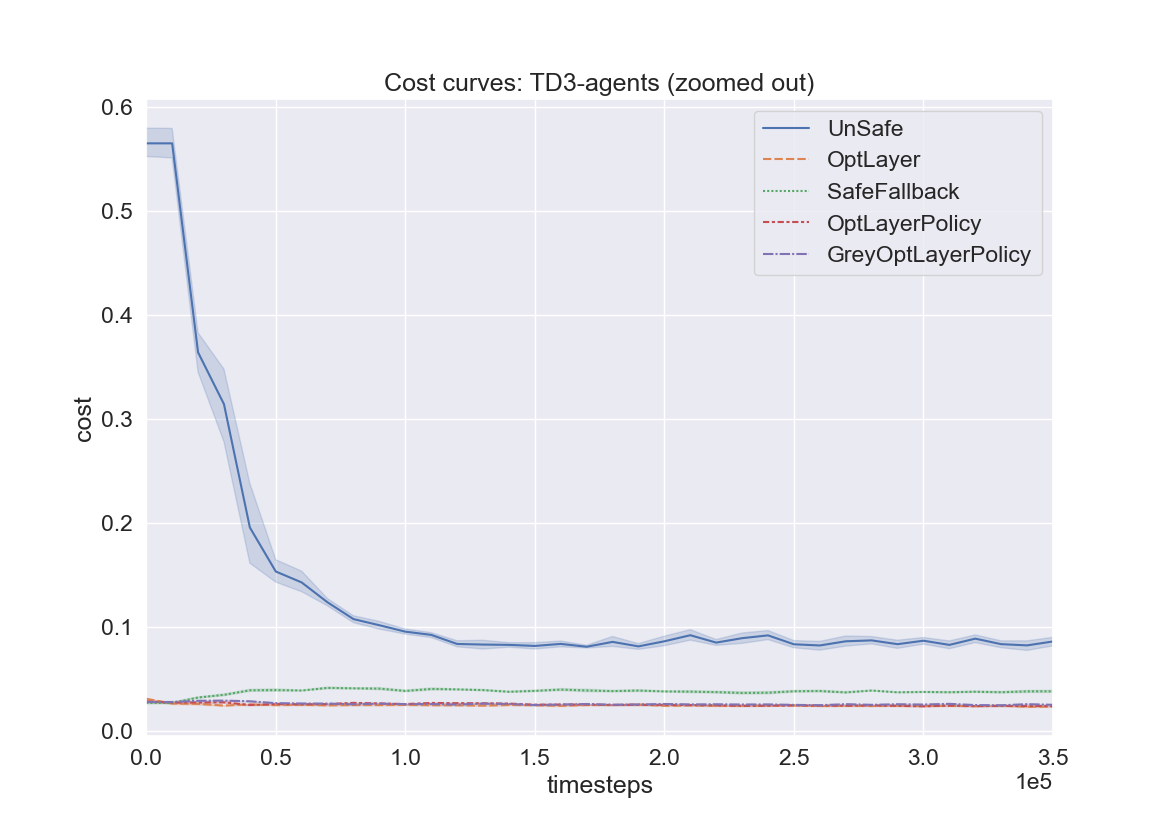}
    \caption{5-run average cost curve (i.e., normalised mean absolute error) with a training budget of 10 years worth of time steps per run (i.e., 350,400 time steps per run). Note that the zoomed-in version is \autoref{fig: cost curve}.}
    \label{fig: cost curve zoomed out}
\end{figure}

\section{Pseudocode and hyperparameters of TD3}
\label{Appendix C}
\scalebox{0.9}{
\begin{algorithm}[H]
\DontPrintSemicolon
\SetAlgoLined
 \nl Input: initial policy parameters \( \theta \), Q-function parameters \( \phi_1 \), \( \phi_2 \), \\ 
 empty replay buffer \(\mathcal{D}\) \;
 \nl Set target parameters equal to main parameters \( \theta_{targ} \leftarrow \theta\), \( \theta_{targ,1} \leftarrow \theta_1\), \( \theta_{targ,2} \leftarrow \theta_2\) \;
 \nl \textbf{repeat} \;
 \nl \hspace{0.2cm} Observe state \(s\) and select action \(a = \text{clip}(\mu_{\theta}(s) + \epsilon, a_{Low}, a_{High}) \), where \( \epsilon \sim \mathcal{N} \) \;
 \nl \hspace{0.2cm} Execute \( a \) in the environment \;
 \nl \hspace{0.2cm} Observe next state \( s' \), reward \( r \) and done signal \( d \) to indicate whether \( s' \) is terminal \;
 \nl \hspace{0.2cm} Store \( (s,a,r,s',d) \) in replay buffer \( \mathcal{D} \) \;
 \nl \hspace{0.2cm} If \( s' \) is terminal, reset environment state \;
 \nl \hspace{0.2cm} \textbf{if} it's time to update \textbf{then} \;
 \nl \hspace{0.4cm} \textbf{for} \( j \) in range(however many updates) \textbf{do} \;
 \nl \hspace{0.6cm} Randomly sample a batch of transitions, \( B = \{ (s,a,r,s',d) \} \) from \( \mathcal{D} \) \;
 \nl \hspace{0.6cm} Compute target actions
 \begin{align*}
     a'(s') = \text{clip}\big( \mu_{\theta_{targ}}(s') + \text{clip}(\epsilon, -c, c), a_{Low}, a_{High} \big),& & \epsilon \sim \mathcal{N}(0,\sigma)
 \end{align*} \vspace{-0.5cm}\;
 \nl \hspace{0.6cm} Compute targets
 \begin{equation*}
     y(r,s',d) = r + \gamma(1 - d) \min_{i=1,2} Q_{\phi_{targ,i}}(s',a'(s'))
 \end{equation*} \vspace{-0.5cm}\;
 \nl \hspace{0.6cm} Update Q-function by one step of gradient descent using
 \begin{align*}
     \nabla_{\phi_{i}} \frac{1}{|B|} \sum_{(s,a,r,s',d)\in B}(Q_{\phi_{i}}(s,a) - y(r,s',d))^2& & \text{for} \ \ i=1,2
 \end{align*} \vspace{-0.5cm}\;
 \nl \hspace{0.6cm} \textbf{if} \( \ j \ \) mod \(\ \texttt{policy\_delay} = 0\) \textbf{then} \;
 \nl \hspace{0.8cm} Update policy by one step of gradient ascent using
 \begin{equation*}
     \nabla_\theta \frac{1}{|B|} \sum_{s\in B} Q_{\phi_{1}}(s,\mu_\theta(s))
 \end{equation*}  \vspace{-0.5cm}\;
 \nl \hspace{0.8cm} Update target networks with
 \begin{align*}
     \phi_{targ,i} \leftarrow \rho \phi_{targ,i} + (1 - \rho) \phi_i& & \text{for} \ \ i=1,2 \\
     \theta_{targ} \leftarrow \rho \theta_{targ} + (1 - \rho) \theta
 \end{align*} \vspace{-0.5cm} \;
 \nl \hspace{0.6cm} \textbf{end if} \;
 \nl \hspace{0.4cm} \textbf{end for} \;
 \nl \hspace{0.2cm} \textbf{end if} \;
 \nl \textbf{until} convergence
 \caption{Twin Delayed DDPG (TD3) \cite{OpenAI2020TwinDocumentation}}\label{algo: td3}
\end{algorithm}
}

\begin{table}[H]
    \centering
    \begin{tabular}{l|c|c}
    \rowcolor[HTML]{efefef} 
    \textbf{Hyperparameters: TD3} & \textbf{Unsafe} & \multicolumn{1}{c}{\begin{tabular}[c]{@{}c@{}}\textbf{OptLayer}\\ \textbf{OptLayerPolicy}\\ \textbf{GreyOptLayerPolicy}\end{tabular}} \\
    \hline
    gamma           &  0.9 & 0.7 \\
    learning\_rate  &  0.0003833 & 0.000583 \\
    batch\_size     &  100 & 16 \\
    buffer\_size    &  1e5 & 1e6 \\
    train\_freq     &  2e3 & 1e0 \\
    gradient\_steps &  2e3 & 1e0 \\
    noise\_type     &  normal & normal \\
    noise\_std      &  0.329 & 0.183 
    \end{tabular}
    \caption{TD3 hyperparameters. The parameters from the \textit{vanilla} Unsafe agent are the result of an optimisation study from \citeauthor{Ceusters2021Model-predictiveStudies} \cite{Ceusters2021Model-predictiveStudies}, while the others are the resulting \texttt{SafeFallback} parameters from \citeauthor{Ceusters2023SafeFunctions} \cite{Ceusters2023SafeFunctions}}
\end{table}

\section{Run-time statistics}
\label{Appendix D}
\setcounter{table}{0}   

\par The experiments are performed on a local machine that has an Intel® Core™ i5-8365U CPU  @1.6GHz, 16 GB of RAM, and an SSD. Over a yearly simulation, the following run-time statistics per simulated time step (with a control horizon of 15 min) are observed.

\begin{table}[!h]
    \centering
    \begin{tabular}{l|c|c|c|c|c}
    \rowcolor[HTML]{efefef} 
    \textbf{Optimal controller} & \textbf{min} & \textbf{mean} & \textbf{std} & \textbf{max} & \textbf{total}\\
    \hline
    Unsafe TD3 & 0,032 s & 0,053 s & 0,009 s & 0,270 s & 1.858 s \\
    Unsafe Random & 0,029 s & 0,046 s & 0,008 s & 0,124 s & 1.630 s \\
    OptLayer TD3 & 0,211 s & 0,291 s & 0,036 s & 0,976 s & 10.204 s \\
    OptLayer Random & 0,177 s & 0,254 s & 0,042 s & 2,256 s & 8.917 s \\
    SafeFallback TD3 & 0,042 s & 0,061 s & 0,007 s & 0,189 s & 2.134 s \\ 
    SafeFallback Random & 0,038 s & 0,046 s & 0,007 s & 0,492 s & 1.608 s \\
    SafeFallback (\(\pi^{safe}\)) & 0,038 s & 0,050 s & 0,010 s & 0,190 s & 1.738 s \\
    OptLayerPolicy TD3 & 0,202 s & 0,257 s & 0,023 s & 1,259 s & 9.008 s \\
    OptLayerPolicy Random & 0,177 s & 0,222 s & 0,028 s & 2,897 s & 7.783 s \\
    GreyOptLayerPolicy TD3 & 1,442 s & 1,944 s & 0,378 s & 10,066 s & 68.106 s \\
    GreyOptLayerPolicy Random & 1,436 s & 1,918 s & 0,361 s & 6,504 s & 67.222 s
    \end{tabular}
    \caption{Run-time statistics \textit{after} training (i.e., pure policy execution)}
    \label{tab: run-time}
\end{table}

\par The maximum run-time per time step never exceeds the control horizon of 15 minutes, as the deployment of the algorithm would then otherwise be considered infeasible with the given hardware. We observe that both the \texttt{UnSafe} and \texttt{SafeFallback} approaches have the fastest run-time, as they do not have a mathematical program to solve. However, we have argued that using a \textit{vanilla} RL agent (i.e., without any safety measures) is not realistic in safety-critical environments and is given here only for completeness. We have also argued that the \texttt{SafeFallback} approach is not capable of handling equality constraints, resulting in a higher constraint tolerance and is seen as a major drawback of the original method. As expected, the run-time of the \texttt{OptLayer} and \texttt{OptLayerPolicy} approaches are in the same order of magnitude, as both involve solving a mixed-integer quadratic program (MIQP) in order to compute the closest feasible action and the distance of the predicted action, \(\tilde a\), to the feasible solution space. Finally, we observe that the \texttt{GreyOptLayerPolicy} approach has the slowest run-time, which in turn hurts its scalability. This is because the surrogate functions that are used to learn the uncertain constraint components are converted within the gradient descent optimisation framework of \texttt{GEKKO} to allow for an exact solution. As proposed in future work, other methods of integrating surrogate functions should be explored to reduce computational complexity.


\reftitle{References}

%=====================================
% References, variant B: external bibliography
%=====================================
%\nocite{*}
\externalbibliography{yes}
\bibliography{references}

\begin{thebibliography}{-------}
\providecommand{\natexlab}[1]{#1}

\bibitem[Fabrizio \em{et~al.}(2009)Fabrizio, Filippi, and Virgone]{Fabrizio2009Trade-offSystems}
Fabrizio, E.; Filippi, M.; Virgone, J.
\newblock {Trade-off between environmental and economic objectives in the optimization of multi-energy systems}.
\newblock {\em Building Simulation 2009 2:1} {\bf 2009}, {\em 2},~29--40.
\newblock doi:{\changeurlcolor{black}\href{https://doi.org/10.1007/S12273-009-9202-4}{\detokenize{10.1007/S12273-009-9202-4}}}.

\bibitem[Engell(2007)]{Engell2007FeedbackOperation}
Engell, S.
\newblock {Feedback control for optimal process operation}.
\newblock {\em Journal of Process Control} {\bf 2007}, {\em 17},~203--219.
\newblock doi:{\changeurlcolor{black}\href{https://doi.org/10.1016/J.JPROCONT.2006.10.011}{\detokenize{10.1016/J.JPROCONT.2006.10.011}}}.

\bibitem[G{\"{o}}rges(2017)]{Gorges2017RelationsLearning}
G{\"{o}}rges, D.
\newblock {Relations between Model Predictive Control and Reinforcement Learning}.
\newblock {\em IFAC-PapersOnLine} {\bf 2017}, {\em 50},~4920--4928.
\newblock doi:{\changeurlcolor{black}\href{https://doi.org/10.1016/j.ifacol.2017.08.747}{\detokenize{10.1016/j.ifacol.2017.08.747}}}.

\bibitem[Ceusters \em{et~al.}(2021)Ceusters, Rodr{\'{i}}guez, Garc{\'{i}}a, Franke, Deconinck, Helsen, Now{\'{e}}, Messagie, and Camargo]{Ceusters2021Model-predictiveStudies}
Ceusters, G.; Rodr{\'{i}}guez, R.C.; Garc{\'{i}}a, A.B.; Franke, R.; Deconinck, G.; Helsen, L.; Now{\'{e}}, A.; Messagie, M.; Camargo, L.R.
\newblock {Model-predictive control and reinforcement learning in multi-energy system case studies}.
\newblock {\em Applied Energy} {\bf 2021}, {\em 303},~117634.
\newblock doi:{\changeurlcolor{black}\href{https://doi.org/10.1016/j.apenergy.2021.117634}{\detokenize{10.1016/j.apenergy.2021.117634}}}.

\bibitem[Ceusters \em{et~al.}(2023)Ceusters, Camargo, Franke, Now{\'{e}}, and Messagie]{Ceusters2023SafeFunctions}
Ceusters, G.; Camargo, L.R.; Franke, R.; Now{\'{e}}, A.; Messagie, M.
\newblock {Safe reinforcement learning for multi-energy management systems with known constraint functions}.
\newblock {\em Energy and AI} {\bf 2023}, {\em 12},~100227.
\newblock doi:{\changeurlcolor{black}\href{https://doi.org/10.1016/J.EGYAI.2022.100227}{\detokenize{10.1016/J.EGYAI.2022.100227}}}.

\bibitem[Pham \em{et~al.}(2018)Pham, De~Magistris, and Tachibana]{Pham2018OptLayerWorld}
Pham, T.H.; De~Magistris, G.; Tachibana, R.
\newblock {OptLayer - Practical Constrained Optimization for Deep Reinforcement Learning in the Real World}.
\newblock {\em Proceedings - IEEE International Conference on Robotics and Automation} {\bf 2018}, pp. 6236--6243.
\newblock doi:{\changeurlcolor{black}\href{https://doi.org/10.1109/ICRA.2018.8460547}{\detokenize{10.1109/ICRA.2018.8460547}}}.

\bibitem[Cao \em{et~al.}(2020)Cao, Hu, Zhao, Zhang, Zhang, Liu, Chen, and Blaabjerg]{Cao2020ReinforcementReview}
Cao, D.; Hu, W.; Zhao, J.; Zhang, G.; Zhang, B.; Liu, Z.; Chen, Z.; Blaabjerg, F.
\newblock {Reinforcement Learning and Its Applications in Modern Power and Energy Systems: A Review}.
\newblock {\em Journal of Modern Power Systems and Clean Energy} {\bf 2020}, {\em 8},~1029--1042.
\newblock doi:{\changeurlcolor{black}\href{https://doi.org/10.35833/MPCE.2020.000552}{\detokenize{10.35833/MPCE.2020.000552}}}.

\bibitem[Yang \em{et~al.}(2020)Yang, Zhao, Li, and Zomaya]{Yang2020ReinforcementSurvey}
Yang, T.; Zhao, L.; Li, W.; Zomaya, A.Y.
\newblock {Reinforcement learning in sustainable energy and electric systems: a survey}.
\newblock {\em Annual Reviews in Control} {\bf 2020}, {\em 49},~145--163.
\newblock doi:{\changeurlcolor{black}\href{https://doi.org/10.1016/J.ARCONTROL.2020.03.001}{\detokenize{10.1016/J.ARCONTROL.2020.03.001}}}.

\bibitem[Perera and Kamalaruban(2021)]{Perera2021ApplicationsSystems}
Perera, A.T.; Kamalaruban, P.
\newblock {Applications of reinforcement learning in energy systems}.
\newblock {\em Renewable and Sustainable Energy Reviews} {\bf 2021}, {\em 137},~110618.
\newblock doi:{\changeurlcolor{black}\href{https://doi.org/10.1016/J.RSER.2020.110618}{\detokenize{10.1016/J.RSER.2020.110618}}}.

\bibitem[Zhou(2022)]{Zhou2022AdvancesPerspectives}
Zhou, Y.
\newblock {Advances of machine learning in multi-energy district communities‒ mechanisms, applications and perspectives}.
\newblock {\em Energy and AI} {\bf 2022}, {\em 10},~100187.
\newblock doi:{\changeurlcolor{black}\href{https://doi.org/10.1016/J.EGYAI.2022.100187}{\detokenize{10.1016/J.EGYAI.2022.100187}}}.

\bibitem[Petrusev \em{et~al.}(2023)Petrusev, Putratama, Rigo-Mariani, Debusschere, Reignier, and Hadjsaid]{Petrusev2023ReinforcementUncertainties}
Petrusev, A.; Putratama, M.A.; Rigo-Mariani, R.; Debusschere, V.; Reignier, P.; Hadjsaid, N.
\newblock {Reinforcement learning for robust voltage control in distribution grids under uncertainties}.
\newblock {\em Sustainable Energy, Grids and Networks} {\bf 2023}, {\em 33},~100959.
\newblock doi:{\changeurlcolor{black}\href{https://doi.org/10.1016/J.SEGAN.2022.100959}{\detokenize{10.1016/J.SEGAN.2022.100959}}}.

\bibitem[Zhou \em{et~al.}(2022)Zhou, Ma, Zhang, and Zou]{Zhou2022Data-drivenLearning}
Zhou, Y.; Ma, Z.; Zhang, J.; Zou, S.
\newblock {Data-driven stochastic energy management of multi energy system using deep reinforcement learning}.
\newblock {\em Energy} {\bf 2022}, {\em 261},~125187.
\newblock doi:{\changeurlcolor{black}\href{https://doi.org/10.1016/J.ENERGY.2022.125187}{\detokenize{10.1016/J.ENERGY.2022.125187}}}.

\bibitem[Pu \em{et~al.}(2021)Pu, Wang, Yang, Yao, and Li]{Pu2021DecomposedLearning}
Pu, Y.; Wang, S.; Yang, R.; Yao, X.; Li, B.
\newblock {Decomposed Soft Actor-Critic Method for Cooperative Multi-Agent Reinforcement Learning}.
\newblock {\em arXiv} {\bf 2021}.

\bibitem[Zhu \em{et~al.}(2022)Zhu, Yang, Liu, Wang, Ma, and Guan]{Zhu2022EnergyPark}
Zhu, D.; Yang, B.; Liu, Y.; Wang, Z.; Ma, K.; Guan, X.
\newblock {Energy Management Based on Multi-Agent Deep Reinforcement Learning for A Multi-Energy Industrial Park}.
\newblock {\em Applied Energy} {\bf 2022}, {\em 311},~118636.
\newblock doi:{\changeurlcolor{black}\href{https://doi.org/10.1016/j.apenergy.2022.118636}{\detokenize{10.1016/j.apenergy.2022.118636}}}.

\bibitem[Ahrarinouri \em{et~al.}(2022)Ahrarinouri, Rastegar, Karami, and Seifi]{Ahrarinouri2022DistributedHubs}
Ahrarinouri, M.; Rastegar, M.; Karami, K.; Seifi, A.R.
\newblock {Distributed reinforcement learning energy management approach in multiple residential energy hubs}.
\newblock {\em Sustainable Energy, Grids and Networks} {\bf 2022}, {\em 32},~100795.
\newblock doi:{\changeurlcolor{black}\href{https://doi.org/10.1016/J.SEGAN.2022.100795}{\detokenize{10.1016/J.SEGAN.2022.100795}}}.

\bibitem[Jendoubi and Bouffard(2022)]{Jendoubi2022Data-drivenLearning}
Jendoubi, I.; Bouffard, F.
\newblock {Data-driven sustainable distributed energy resources’ control based on multi-agent deep reinforcement learning}.
\newblock {\em Sustainable Energy, Grids and Networks} {\bf 2022}, {\em 32},~100919.
\newblock doi:{\changeurlcolor{black}\href{https://doi.org/10.1016/J.SEGAN.2022.100919}{\detokenize{10.1016/J.SEGAN.2022.100919}}}.

\bibitem[Sun \em{et~al.}(2023)Sun, Song, Li, Zou, Pan, Lu, Yang, Zhang, and Kong]{Sun2023Multi-objectiveAlgorithm}
Sun, B.; Song, M.; Li, A.; Zou, N.; Pan, P.; Lu, X.; Yang, Q.; Zhang, H.; Kong, X.
\newblock {Multi-objective solution of optimal power flow based on TD3 deep reinforcement learning algorithm}.
\newblock {\em Sustainable Energy, Grids and Networks} {\bf 2023}, {\em 34},~101054.
\newblock doi:{\changeurlcolor{black}\href{https://doi.org/10.1016/J.SEGAN.2023.101054}{\detokenize{10.1016/J.SEGAN.2023.101054}}}.

\bibitem[Garc{\'{i}}a and Fern{\'{a}}ndez(2015)]{Garcia2015ALearning}
Garc{\'{i}}a, J.; Fern{\'{a}}ndez, F.
\newblock {A comprehensive survey on safe reinforcement learning},  2015.

\bibitem[Feng \em{et~al.}(2023)Feng, Wang, Yang, Chen, Li, Yang, and Wang]{Feng2023EconomicApproach}
Feng, J.; Wang, H.; Yang, Z.; Chen, Z.; Li, Y.; Yang, J.; Wang, K.
\newblock {Economic dispatch of industrial park considering uncertainty of renewable energy based on a deep reinforcement learning approach}.
\newblock {\em Sustainable Energy, Grids and Networks} {\bf 2023}, {\em 34},~101050.
\newblock doi:{\changeurlcolor{black}\href{https://doi.org/10.1016/J.SEGAN.2023.101050}{\detokenize{10.1016/J.SEGAN.2023.101050}}}.

\bibitem[Zhu \em{et~al.}(2020)Zhu, Yang, Liu, Ma, Zhu, Ma, and Guan]{Zhu2020EnergyNetworks}
Zhu, D.; Yang, B.; Liu, Q.; Ma, K.; Zhu, S.; Ma, C.; Guan, X.
\newblock {Energy trading in microgrids for synergies among electricity, hydrogen and heat networks}.
\newblock {\em Applied Energy} {\bf 2020}, {\em 272},~115225.
\newblock doi:{\changeurlcolor{black}\href{https://doi.org/10.1016/J.APENERGY.2020.115225}{\detokenize{10.1016/J.APENERGY.2020.115225}}}.

\bibitem[Zhu \em{et~al.}(2022)Zhu, Yang, Ma, Wang, Zhu, Ma, and Guan]{Zhu2022StochasticConstraints}
Zhu, D.; Yang, B.; Ma, C.; Wang, Z.; Zhu, S.; Ma, K.; Guan, X.
\newblock {Stochastic gradient-based fast distributed multi-energy management for an industrial park with temporally-coupled constraints}.
\newblock {\em Applied Energy} {\bf 2022}, {\em 317},~119107.
\newblock doi:{\changeurlcolor{black}\href{https://doi.org/10.1016/J.APENERGY.2022.119107}{\detokenize{10.1016/J.APENERGY.2022.119107}}}.

\bibitem[Zou \em{et~al.}(2023{\natexlab{a}})Zou, Xu, Feng, and Nguyen]{Zou2023Peer-to-PeerNetwork}
Zou, Y.; Xu, Y.; Feng, X.; Nguyen, H.D.
\newblock {Peer-to-Peer Transactive Energy Trading of a Reconfigurable Multi-Energy Network}.
\newblock {\em IEEE Transactions on Smart Grid} {\bf 2023}, {\em 14},~2236--2249.
\newblock doi:{\changeurlcolor{black}\href{https://doi.org/10.1109/TSG.2022.3223378}{\detokenize{10.1109/TSG.2022.3223378}}}.

\bibitem[Zou \em{et~al.}(2023{\natexlab{b}})Zou, Xu, and Zhang]{Zou2023AMicrogrid}
Zou, Y.; Xu, Y.; Zhang, C.
\newblock {A Risk-Averse Adaptive Stochastic Optimization Method for Transactive Energy Management of a Multi-Energy Microgrid}.
\newblock {\em IEEE Transactions on Sustainable Energy} {\bf 2023}, {\em 14},~1599--1611.
\newblock doi:{\changeurlcolor{black}\href{https://doi.org/10.1109/TSTE.2023.3240184}{\detokenize{10.1109/TSTE.2023.3240184}}}.

\bibitem[Brunke \em{et~al.}(2021)Brunke, Greeff, Hall, Yuan, Zhou, Panerati, and Schoellig]{Brunke2021SafeLearning}
Brunke, L.; Greeff, M.; Hall, A.W.; Yuan, Z.; Zhou, S.; Panerati, J.; Schoellig, A.P.
\newblock {Safe Learning in Robotics: From Learning-Based Control to Safe Reinforcement Learning}.
\newblock {\em Annual Review of Control, Robotics, and Autonomous Systems} {\bf 2021}, {\em 5}.
\newblock doi:{\changeurlcolor{black}\href{https://doi.org/10.1146/annurev-control-042920-020211}{\detokenize{10.1146/annurev-control-042920-020211}}}.

\bibitem[McKinnon and Schoellig(2020)]{McKinnon2020Context-awareControl}
McKinnon, C.D.; Schoellig, A.P.
\newblock {Context-aware Cost Shaping to Reduce the Impact of Model Error in Receding Horizon Control}.
\newblock {\em Proceedings - IEEE International Conference on Robotics and Automation} {\bf 2020}, pp. 2386--2392.
\newblock doi:{\changeurlcolor{black}\href{https://doi.org/10.1109/ICRA40945.2020.9197521}{\detokenize{10.1109/ICRA40945.2020.9197521}}}.

\bibitem[Bharadhwaj \em{et~al.}(2021)Bharadhwaj, Kumar, Rhinehart, Levine, Shkurti, and Garg]{Bharadhwaj2021ConservativeExploration}
Bharadhwaj, H.; Kumar, A.; Rhinehart, N.; Levine, S.; Shkurti, F.; Garg, A.
\newblock {Conservative Safety Critics for Exploration}.
\newblock  International Conference on Learning Representations,  2021.

\bibitem[Lopez \em{et~al.}(2021)Lopez, Slotine, and How]{Lopez2021RobustSafety}
Lopez, B.T.; Slotine, J.J.E.; How, J.P.
\newblock {Robust Adaptive Control Barrier Functions: An Adaptive and Data-Driven Approach to Safety}.
\newblock {\em IEEE Control Systems Letters} {\bf 2021}, {\em 5},~1031--1036.
\newblock doi:{\changeurlcolor{black}\href{https://doi.org/10.1109/LCSYS.2020.3005923}{\detokenize{10.1109/LCSYS.2020.3005923}}}.

\bibitem[Sutton and Barto(2018)]{Sutton2018ReinforcementIntroduction}
Sutton, R.S.; Barto, A.G.
\newblock {\em {Reinforcement learning: An introduction}}; MIT press,  2018.

\bibitem[Gunnell \em{et~al.}(2022)Gunnell, Manwaring, Lu, Reynolds, Vienna, and Hedengren]{Gunnell2022MachineConstraints}
Gunnell, L.L.; Manwaring, K.; Lu, X.; Reynolds, J.; Vienna, J.; Hedengren, J.
\newblock {Machine Learning with Gradient-Based Optimization of Nuclear Waste Vitrification with Uncertainties and Constraints}.
\newblock {\em Processes} {\bf 2022}, {\em 10},~2365.
\newblock doi:{\changeurlcolor{black}\href{https://doi.org/10.3390/pr10112365}{\detokenize{10.3390/pr10112365}}}.

\bibitem[Beal \em{et~al.}(2018)Beal, Hill, Martin, and Hedengren]{Beal2018GEKKOSuite}
Beal, L.; Hill, D.; Martin, R.; Hedengren, J.
\newblock {GEKKO Optimization Suite}.
\newblock {\em Processes} {\bf 2018}, {\em 6},~106.
\newblock doi:{\changeurlcolor{black}\href{https://doi.org/10.3390/pr6080106}{\detokenize{10.3390/pr6080106}}}.

\bibitem[Mattsson \em{et~al.}(1998)Mattsson, Elmqvist, and Otter]{Mattsson1998PhysicalModelica}
Mattsson, S.E.; Elmqvist, H.; Otter, M.
\newblock {Physical system modeling with Modelica}.
\newblock  Control Engineering Practice. Pergamon,  1998, Vol.~6, pp. 501--510.
\newblock doi:{\changeurlcolor{black}\href{https://doi.org/10.1016/S0967-0661(98)00047-1}{\detokenize{10.1016/S0967-0661(98)00047-1}}}.

\bibitem[Gr{\"{a}}ber \em{et~al.}(2017)Gr{\"{a}}ber, Fritzsche, and Tegethoff]{Graber2017FromProblems}
Gr{\"{a}}ber, M.; Fritzsche, J.; Tegethoff, W.
\newblock {From system model to optimal control - A tool chain for the efficient solution of optimal control problems}.
\newblock  Proceedings of the 12th International Modelica Conference, Prague, Czech Republic, May 15-17, 2017. Link{\"{o}}ping University Electronic Press,  2017, Vol. 132, pp. 249--254.
\newblock doi:{\changeurlcolor{black}\href{https://doi.org/10.3384/ecp17132249}{\detokenize{10.3384/ecp17132249}}}.

\bibitem[Brockman \em{et~al.}(2016)Brockman, Cheung, Pettersson, Schneider, Schulman, Tang, and Zaremba]{Brockman2016OpenAIGym}
Brockman, G.; Cheung, V.; Pettersson, L.; Schneider, J.; Schulman, J.; Tang, J.; Zaremba, W.
\newblock {OpenAI Gym}.
\newblock {\em arxiv} {\bf 2016}.

\bibitem[Pedregosa~FABIANPEDREGOSA \em{et~al.}(2011)Pedregosa~FABIANPEDREGOSA, Michel, Grisel~OLIVIERGRISEL, Blondel, Prettenhofer, Weiss, Vanderplas, Cournapeau, Pedregosa, Varoquaux, Gramfort, Thirion, Grisel, Dubourg, Passos, Brucher, Perrot~and{\'{E}}douardand, Duchesnay, and Duchesnay~EDOUARDDUCHESNAY]{PedregosaFABIANPEDREGOSA2011Scikit-learn:Python}
Pedregosa~FABIANPEDREGOSA, F.; Michel, V.; Grisel~OLIVIERGRISEL, O.; Blondel, M.; Prettenhofer, P.; Weiss, R.; Vanderplas, J.; Cournapeau, D.; Pedregosa, F.; Varoquaux, G.; Gramfort, A.; Thirion, B.; Grisel, O.; Dubourg, V.; Passos, A.; Brucher, M.; Perrot~and{\'{E}}douardand, M.; Duchesnay, a.; Duchesnay~EDOUARDDUCHESNAY, F.
\newblock {Scikit-learn: Machine learning in Python}.
\newblock {\em jmlr.orgF Pedregosa, G Varoquaux, A Gramfort, V Michel, B Thirion, O Grisel, M Blondelthe Journal of machine Learning research, 2011•jmlr.org} {\bf 2011}, {\em 12},~2825--2830.

\bibitem[Andersson \em{et~al.}(2016)Andersson, Akesson, and Fuhrer]{Andersson2016PyFMI:Interface}
Andersson, C.; Akesson, J.; Fuhrer, C.
\newblock {PyFMI: A Python Package for Simulation of Coupled Dynamic Models with the Functional Mock-up Interface}.
\newblock Technical Report~2, Lund University,  2016.

\bibitem[Drgoňa \em{et~al.}(2020)Drgoňa, Arroyo, Cupeiro~Figueroa, Blum, Arendt, Kim, Oll{\'{e}}, Oravec, Wetter, Vrabie, and Helsen]{Drgona2020AllBuildings}
Drgoňa, J.; Arroyo, J.; Cupeiro~Figueroa, I.; Blum, D.; Arendt, K.; Kim, D.; Oll{\'{e}}, E.P.; Oravec, J.; Wetter, M.; Vrabie, D.L.; Helsen, L.
\newblock {All you need to know about model predictive control for buildings}.
\newblock {\em Annual Reviews in Control} {\bf 2020}, {\em 50},~190--232.
\newblock doi:{\changeurlcolor{black}\href{https://doi.org/10.1016/j.arcontrol.2020.09.001}{\detokenize{10.1016/j.arcontrol.2020.09.001}}}.

\bibitem[Kingma and Ba(2014)]{Kingma2014Adam:Optimization}
Kingma, D.P.; Ba, J.
\newblock {Adam: A Method for Stochastic Optimization} {\bf 2014}.

\bibitem[Raffin \em{et~al.}(2021)Raffin, Hill, Gleave, Kanervisto, Ernestus, and Dormann]{stable-baselines3}
Raffin, A.; Hill, A.; Gleave, A.; Kanervisto, A.; Ernestus, M.; Dormann, N.
\newblock {Stable-Baselines3: Reliable Reinforcement Learning Implementations}.
\newblock {\em Journal of Machine Learning Research} {\bf 2021}, {\em 22},~1--8.

\bibitem[{OpenAI}(2020)]{OpenAI2020TwinDocumentation}
{OpenAI}.
\newblock {Twin Delayed DDPG — Spinning Up documentation},  2020.

\end{thebibliography}

%%%%%%%%%%%%%%%%%%%%%%%%%%%%%%%%%%%%%%%%%%
%% optional
%\sampleavailability{Samples of the compounds ...... are available from the authors.}

%\reviewreports{\\
%Reviewer 1 comments and authors’ response\\
%Reviewer 2 comments and authors’ response\\
%Reviewer 3 comments and authors’ response
%}

%%%%%%%%%%%%%%%%%%%%%%%%%%%%%%%%%%%%%%%%%%
\end{document}